\documentclass[preprint,showpacs,preprintnumbers,amsmath,amssymb,aps,floatfix]{revtex4}

\usepackage{dcolumn}% Align table columns on decimal point
\usepackage{graphicx}% Include figure files
\def\LL{\left\langle}	% left angle bracket
\def\RR{\right\rangle}	% right angle bracket
\def\LP{\left(}		% left parenthesis
\def\RP{\right)}	% right parenthesis
\def\PAR#1#2{ {\frac{\partial #1}{\partial #2}} }

\newcommand{\BE}{\begin{displaymath}}
\newcommand{\EE}{\end{displaymath}}
\newcommand{\BNE}{\begin{equation}}
\newcommand{\ENE}{\end{equation}}
\newcommand{\BEA}{\begin{eqnarray}}
\newcommand{\EEA}{\nonumber\end{eqnarray}}
\newcommand{\EL}{\nonumber\\}

\newcommand{\Tr}{{\rm Tr}}
\renewcommand{\Re}{{\rm Re}}
\newcommand{\pbp}{\bar\psi\psi}

\newcommand{\Dslash}{\makebox[0pt][l]{\,/}D}

\begin{document}

\title{Scaling studies of QCD with the dynamical HISQ action}

\author{A. Bazavov}
\affiliation{Department of Physics, University of Arizona, Tucson, AZ 85721, USA}
\author{C. Bernard}
\affiliation{Department of Physics, Washington University, St. Louis, MO 63130, USA}
\author{C. DeTar}
\affiliation{Physics Department, University of Utah, Salt Lake City, UT 84112, USA}
\author{W. Freeman}
\affiliation{Department of Physics, University of Arizona, Tucson, AZ 85721, USA}
\author{Steven Gottlieb}
\affiliation{Department of Physics, Indiana University, Bloomington, IN 47405, USA}
\affiliation{National Center for Supercomputing Applications, University of Illinois, Urbana, IL
61801, USA}
\author{U.M. Heller}
\affiliation{American Physical Society, One Research Road, Ridge, NY 11961, USA}
\author{J.E. Hetrick}
\affiliation{Physics Department, University of the Pacific, Stockton, CA 95211, USA}
\author{J. Laiho
\footnote{Present address: University of Glasgow, Glasgow G12 8QQ, UK}
}
\affiliation{Department of Physics, Washington University, St. Louis, MO 63130, USA}
\author{L. Levkova}
\affiliation{Physics Department, University of Utah, Salt Lake City, UT 84112, USA}
\author{M. Oktay}
\affiliation{Physics Department, University of Utah, Salt Lake City, UT 84112, USA}
\author{J. Osborn}
\affiliation{Argonne Leadership Computing Facility, Argonne National Laboratory, Argonne, IL
60439, USA}
\author{R.L. Sugar}
\affiliation{Department of Physics, University of California, Santa Barbara, CA 93106, USA}
\author{D. Toussaint}
\affiliation{Department of Physics, University of Arizona, Tucson, AZ 85721, USA}
\author{R.S. Van de Water}
\affiliation{Department of Physics, Brookhaven National Laboratory, Upton, NY 11973, USA}
\author{[MILC Collaboration]}
%\collaboration{MILC Collaboration}\noaffiliation

\date{\today}

\begin{abstract}
We study the lattice spacing dependence, or scaling, of physical
quantities using the highly improved staggered quark (HISQ) 
action introduced by the HPQCD/UKQCD collaboration,
comparing our results to similar simulations with the asqtad
fermion action.  Results are based on calculations with lattice spacings approximately
0.15, 0.12 and 0.09 fm, using four flavors of dynamical HISQ quarks.
The strange and charm quark masses are near their physical values,
and the light-quark mass is set to 0.2 times the strange-quark mass.
We look at the lattice spacing dependence of hadron
masses, pseudoscalar meson decay constants, and the topological
susceptibility.  In addition to the commonly used determination of the
lattice spacing through the static quark potential, we examine a
determination proposed by the HPQCD collaboration that uses
the decay constant of a fictitious ``unmixed $s\bar s$'' pseudoscalar
meson. We find that the lattice
artifacts in the HISQ simulations are much smaller than those in
the asqtad simulations at the same lattice spacings and quark masses.
\end{abstract}
\pacs{12.38.Gc,14.20.Dh}

\maketitle

\section{Introduction and motivation}

The ``highly improved staggered quark'', or HISQ, action was developed
by the HPQCD/UKQCD collaboration to reduce the lattice artifacts associated
with staggered quarks in lattice QCD
calculations~\cite{hpqcd_hisq2003,hpqcd_hisq2004,hpqcd_hisqprd}.
While significantly more expensive than the asqtad action used in the MILC
collaboration's long-running program of QCD simulations with three dynamical
quark flavors~\cite{milc_rmp}, it is still very economical compared with non-staggered
quark actions.

The initial studies of the HISQ action by the HPQCD/UKQCD collaboration
demonstrated the reduction of
taste symmetry breaking and improvements in the dispersion relation for
the charm quark by using the HISQ action for valence quarks on quenched
lattices and lattices 
generated with asqtad sea quarks~\cite{hpqcd_hisq2003,hpqcd_hisq2004,hpqcd_hisqprd}.  Further work
with this action, again implemented for the valence quarks with asqtad sea
quarks, has demonstrated impressive precision for charmonium and heavy-light meson
physics~\cite{hpqcd_lat2006,hpqcd_hisqpseudoscalars,hqcd_lat2008}.

As a first stage in a complete program of QCD simulations using the HISQ action
for dynamical quarks, we have generated ensembles of lattices at three different
lattice spacings with four flavors of dynamical quarks, where the light-quark mass 
is fixed at two-tenths of the strange
quark mass, and the strange and charm quark masses are near their 
physical values.  This allows us to test scaling, or dependence of calculated
quantities on the lattice spacing. The purpose of this paper is to
report on these tests at fixed quark mass.   Where possible, we compare
the lattice spacing dependence of physical quantities
with the HISQ action to their dependence using the asqtad action
at the same quark mass and lattice spacings.  We look at the static quark
potential, splittings among the different tastes of pions, masses of
the rho and nucleon,
%the charmonium hyperfine splitting,
pseudoscalar meson decay constants and the topological susceptibility.
 We emphasize that all of this is done at a fixed,
and unphysically large, light-quark mass --- our purpose here is to make
a controlled study of the dependence on lattice spacing.

%\begin{verbatim}
%OUTLINE
%
%Differences vs asqtad
%  smeared links in Dslash
%  correction to Naik term for charm
%  four dynamical flavors
%  fermion loops in gauge action coefficients
%
%Use same analysis as asqtad where practical, since purpose is
%to compare the actions.  Note we will have to interpolate in
%asqtad results to properly match quark masses.
%
%Notes on algorithm used and costs
%
%Tabulate parameters of HISQ runs
%
%What to look at: need dimensionless ratios
%mass * r_1  - commonly used, but much of asqtad scaling violation
%is just lattice spacing dependence of r1
%mass / f_ss, where f_ss is PS decay constant at strange-quark mass
%   unphysical, but can be calculated accurately
%
%Static quark potential
%  commonly used as interpolating quantity in setting lattice spacing
%  we see larger short distance artifacts than with asqtad -- possibly
%    from fermion loop corrections to coefficients?
%
%Taste splittings
%  Same old graph
%
%Rho mass
%  Same old graph, but add m_rho/f_ss
%
%Nucleon mass
%  Same old graph
%
%Charmonium masses and dispersion relation
%
%We need to think about whether r1 is "bad scaling variable".  That is,
%since most masses in units of r1 increase as lattice spacing increases,
%is  it really just a problem with r1?
%We could look at ratios of masses to f_ss, for example.
%Or other dimensionless ratios?
%
%Topological susceptibility
%\end{verbatim}

\section{Methods and lattice data}

There are four major differences between these HISQ simulations and our
earlier asqtad simulations.

 First, the HISQ simulations include the
effects of a dynamical charm quark.  We expect that the effects of dynamical
charm will be very small for the quantities studied here, but with modern
algorithms it is cheap to include the charm quark, and we plan to
investigate quantities involving dynamical charm in the future.

Second, the one-quark-loop contributions to the perturbative calculation
of the coefficients in the Symanzik improved gauge action are included.
At the time the asqtad simulation program was started these corrections
% Steve wants comma after started
were not available, but they have now been computed for both the asqtad and
HISQ actions, and are
unexpectedly large~\cite{GAUGE_COEFFS}.

Third, in the HISQ action the parallel transport of quark fields is
done with a link that is highly smeared.   Specifically, it is first
smeared using a ``fat7'' smearing, then projected onto a unitary
matrix, and then smeared again with an ``asqtad'' smearing~\cite{hpqcd_hisqprd}.
The use of the asqtad smearing in the second iteration, together with
the addition of the Naik term, or third-nearest-neighbor coupling,
in $\Dslash$, insures that the fermion action is formally order $a^2$
improved.   The use of two levels of smearing produces a smooth gauge
field as seen by the quarks, and this explains the reduced taste symmetry
violations.

Finally, the third-nearest-neighbor term in the charm quark $\Dslash$ is
modified to improve the charm quark dispersion relation~\cite{hpqcd_hisqprd}.
These last two differences combine to make up what is usually meant by
``the HISQ action'', although in principle they could be introduced
independently.

Where practical, since our purpose is to compare the lattice artifacts
in the two actions, we use the same analysis for the HISQ data as was used
for the asqtad data.

Table~\ref{tab:hisqruns} shows the parameters of the three HISQ runs
used in these tests.   Detailed information about the asqtad ensembles
can be found in Ref.~\cite{milc_rmp}.

The HISQ lattices were generated using the rational hybrid Monte Carlo (RHMC)
algorithm~\cite{RHMC}.   Issues with implementing this algorithm for
the HISQ action have been discussed in Ref.~\cite{milc_hisq}.
We used different molecular dynamics step sizes for the gauge and fermion parts
of the action, with three gauge steps for each fermion step~\cite{SEXTON}.
We used the Omelyan integration algorithm in both the gauge and fermion parts~\cite{SEXTON,OMELYAN}.
Five pseudofermion fields were used, each with a rational function approximation
for the fractional powers.  The first 
implements the ratio of the roots of the determinants for the light
and strange sea quarks to the determinant for three heavy ``regulator'' quarks with
mass $am_r=0.2$.
That is, it corresponds to the weight $\det \LP M(m_l) \RP^{1/2} \det \LP M(m_s) \RP^{1/4} \det \LP M(m_r) \RP^{-3/4} $.
The next three pseudofermion fields each
implement the force from one flavor of the regulator quark, or the fourth root of
the corresponding determinant~\cite{HASENBUSCH}. The final pseudofermion field
implements the dynamical charm quark.

Rational function approximations were used for the fractional powers
of the matrices \cite{RHMC,CLARKCODE}.  In the molecular dynamics evolution
we used a 9'th order approximation for the pseudofermion field containing
the light quarks, and a 7'th order approximation for the three regulator fields
and the charm quark pseudofermion.   For the heat bath updating of the 
pseudofermion fields and for computing the action at the beginning and end
of the molecular dynamics trajectory we used 11'th order and 9'th order
approximations.   These approximations comfortably exceeded the required
accuracy, but since a multimass conjugate gradient routine is used for the
sparse matrix solutions, adding extra terms in these approximations has minimal
cost.

In order to make this paper self-contained, we summarize the action
in Appendix~\ref{app_HISQ_action}, and discuss some algorithmic issues 
specific to the HISQ action in Appendices~\ref{app_HISQ_force}, \ref{app_algebra},
and~\ref{app_charm}.

% SAY WHAT SMOOTHING PARAMETERS IN ACTION, AND SVD WHEN NEEDED.
% HISQ_FORCE_FILTER=0.000050 HISQ_REUNIT_ALLOW_SVD RESIDUAL_TRICK=1.000000e-12

\begin{center}\begin{table}
\caption{
\label{tab:hisqruns}
Parameters of the HISQ runs with $m_l=0.2\, m_s$.
%Here $\epsilon_N$ is the correction for the three link (Naik) term in the
%charm quark action, and the perturbative expression we use is given
%in Appendix~\ref{app_HISQ_action}.
Here $\epsilon_N$ is the correction for the three link (Naik) term in the
charm quark action.
These values differ slightly from the expression in Appendix~\ref{app_HISQ_action}
because they do not include the distinction between bare and tree-level quark
mass (see Eq.~(24) in Ref.~\cite{hpqcd_hisqprd}.)
The expression in Appendix~\ref{app_HISQ_action} is used in all more
recent ensembles.
The number
of equilibrated lattices is $N_{lats}$. The separation of the lattices
in simulation time is $S_t$, the length of a trajectory in simulation time is $L_t$, the molecular
dynamics step size is $\epsilon$, and the fraction of trajectories accepted is ``acc.''.
Our definition of the step size is such that there is one evaluation of the fermion force
per step, so a complete cycle of the Omelyan integration algorithm includes two
fermion-action steps and six gauge-action steps.
The physical lattice spacing given in this table uses the three
flavor determination of
$r_1=0.3117(6)(\null_{-31}^{+12})$ fm made
using $f_\pi$ to set the scale on the asqtad ensembles~\cite{milc_fpi_lat09}.
It should be noted that when chiral and continuum limits of 2+1+1 flavor
calculations are completed, a 2+1+1 flavor determination of $r_1$ will supercede this.
}
\begin{tabular}{|l|llll|l|l|lllll|ll|}
\hline
$\frac{10}{g^2}$ & $am_l$ & $am_s$ & $am_c$ & $\epsilon_N$ & size	& $u_0$ & 
    $N_{lats}$ & $S_t$ & $L_t$ & $\epsilon$	& acc. & $r_1/a$ & $a$ (fm) \\ \hline
5.8     &  0.013 & 0.065  & 0.838  & $-0.3582$	& $16^3\times 48$	& 0.85535 & 
    1021     & 5    & 1.0	& 0.033	& 0.73	&  2.041(10) & $0.1527(\null_{-16}^{+7})$ \\
6.0     & 0.0102 & 0.0509 & 0.635  & $-0.2308$	& $24^3\times 64$	& 0.86372 & 
    1040     & 5    & 1.0	& 0.036	& 0.66	&  2.574(5)  & $0.1211(\null_{-12}^{+6})$ \\
6.3     & 0.0074 & 0.037  & 0.440  & $-0.1205$	& $32^3\times 96$	& 0.874164& 
    878      & 6    & 1.5	& 0.031	& 0.68	&  3.520(7)  & $0.0886(\null_{-9}^{+4})$ \\
\hline
\end{tabular}
\end{table}\end{center}

\section{Autocorrelations in simulation time}

% TRY THIS
\begin{figure}[t]
\hspace{-0.1in}
\rule{0.0in}{0.0in}\vspace{-1.0in}\\
\begin{tabular}{lll}
\includegraphics[width=0.33\textwidth]{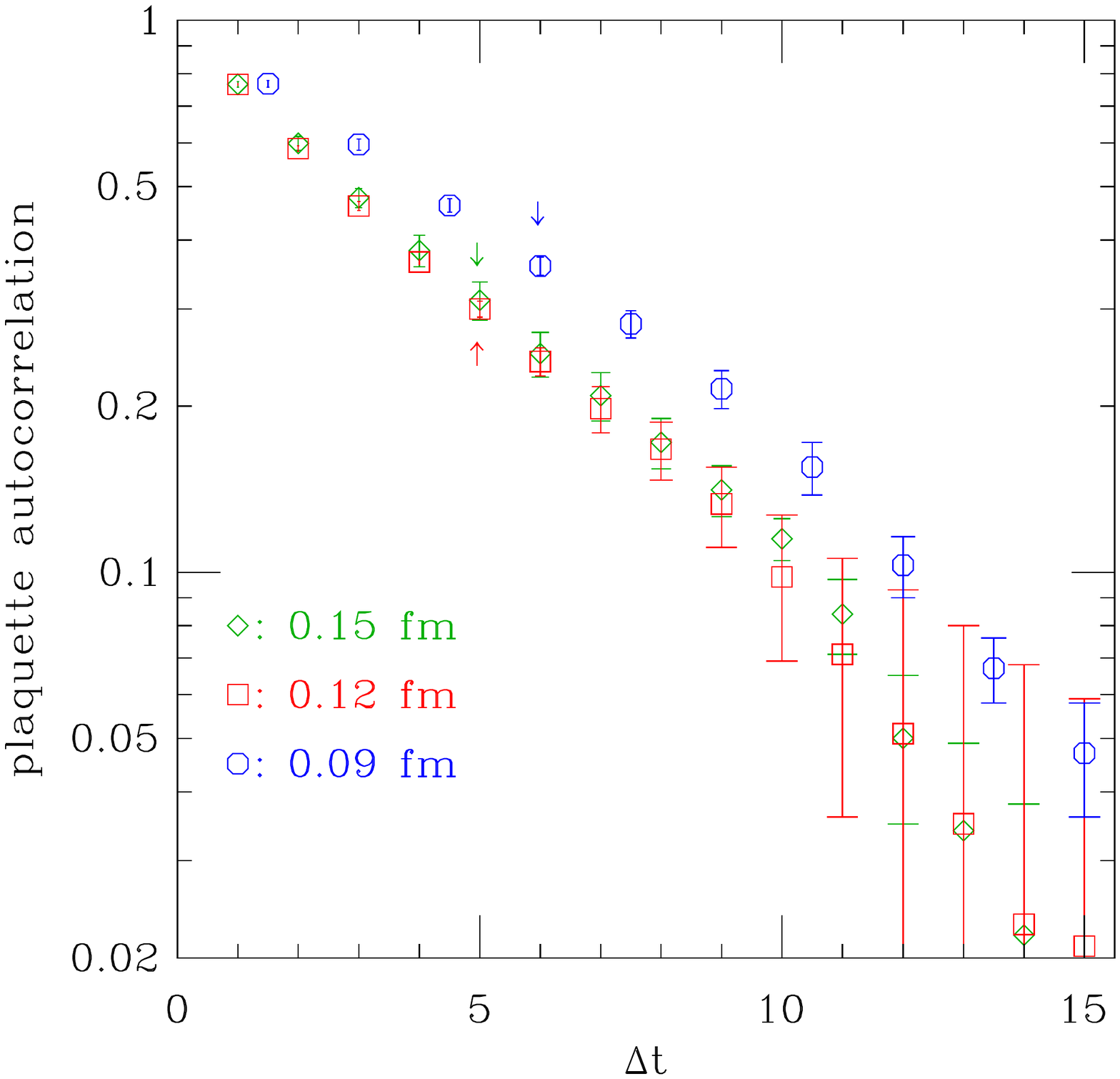} &
\includegraphics[width=0.33\textwidth]{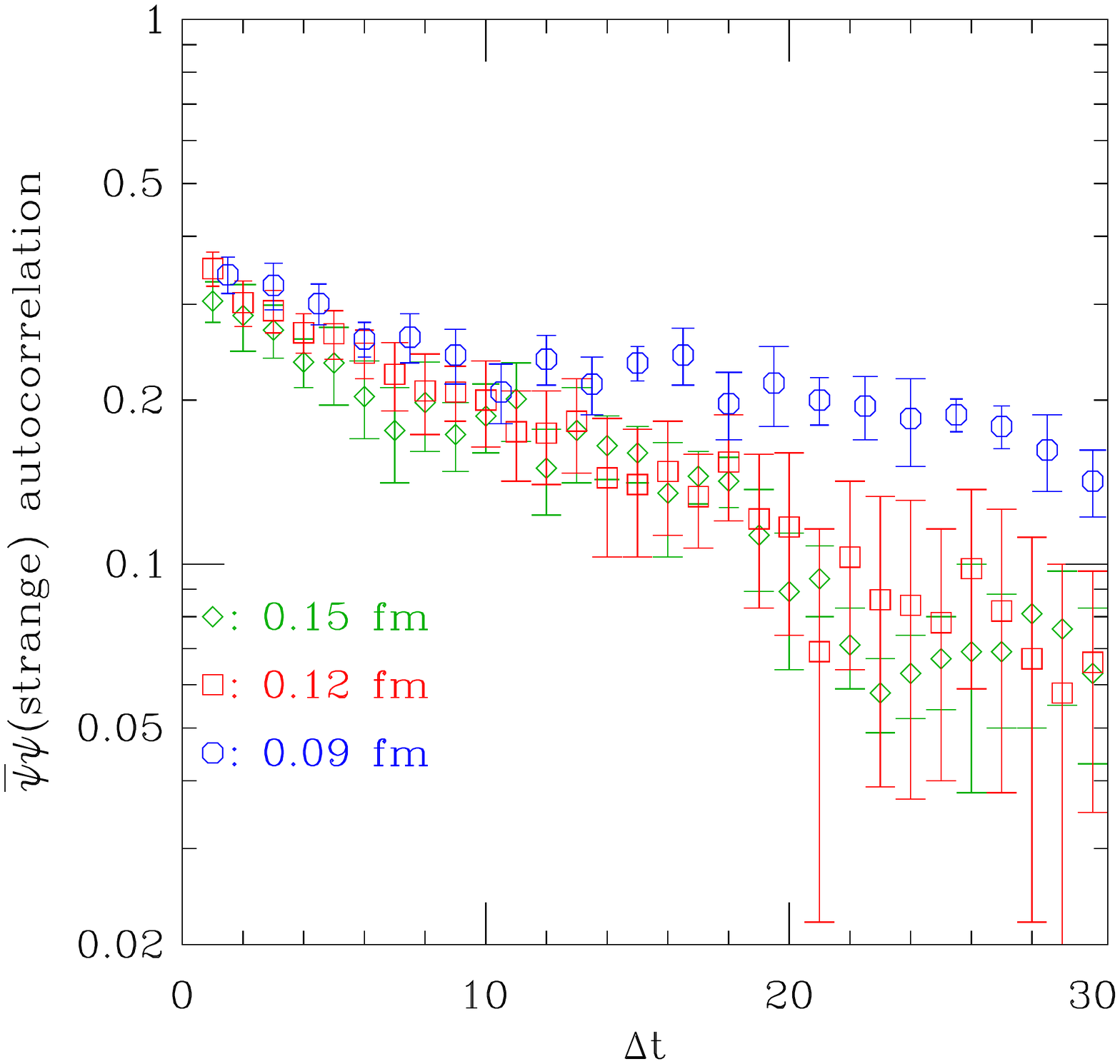} &
\includegraphics[width=0.33\textwidth]{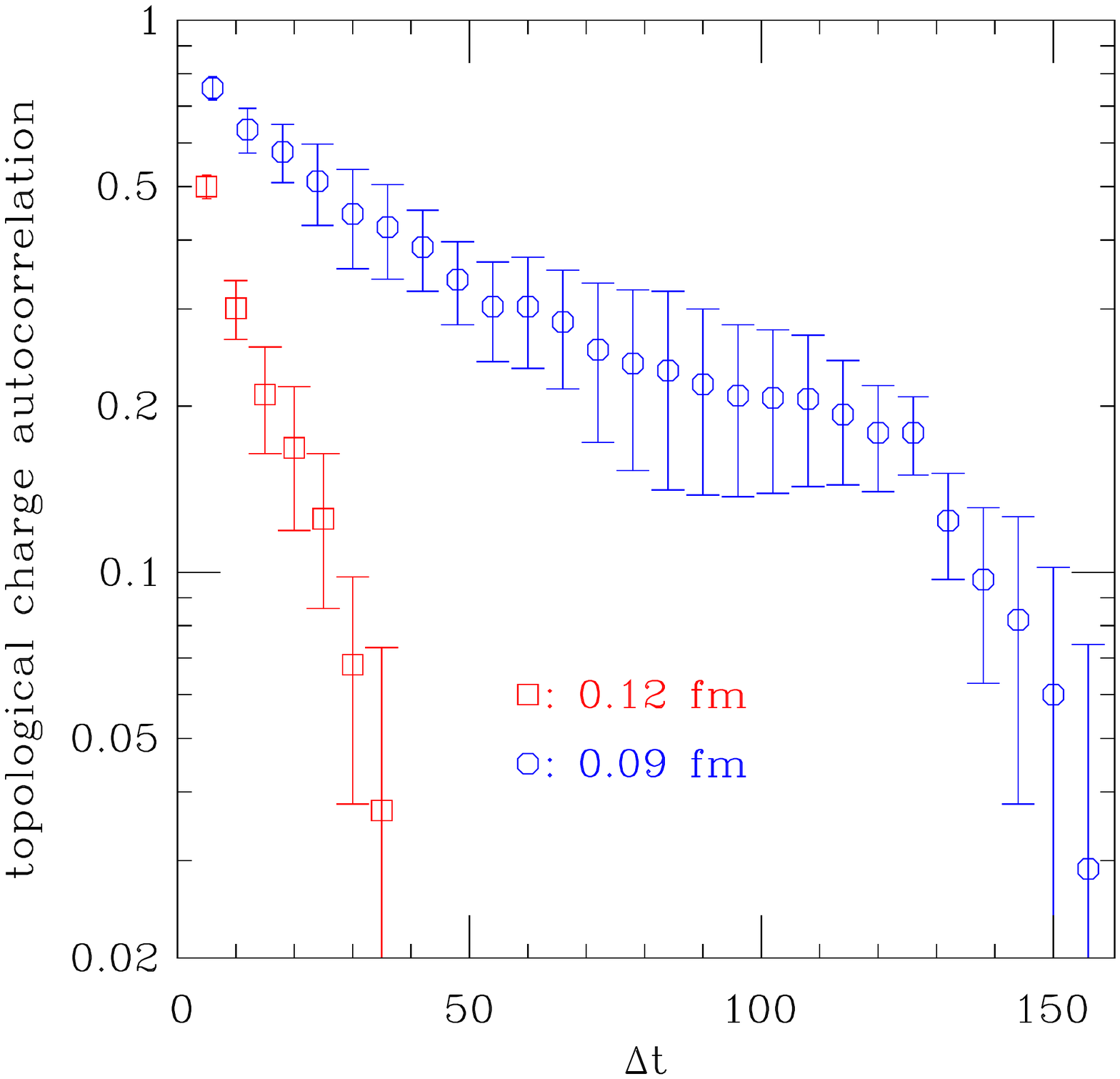} \\
\end{tabular}
\rule{0.0in}{0.0in}\vspace{-0.5in}\\
\caption{
Autocorrelation $C_{\Delta t}$ in simulation time of the plaquette (left panel),
strange quark $\pbp$ (center panel) and topological charge (right panel).
Note that the horizontal scale is different in each of the three panels.
Errors on the autocorrelation were estimated by dividing the time series into five
subsets and averaging the autocorrelations from each subset.
The vertical arrows in the left panel indicate the time separation between stored lattices,
used in computing the potential, spectrum and other quantities.
\label{fig:autocor_all}}
\end{figure}

%\begin{figure}[t]
%\hspace{-0.1in}
%\rule{0.0in}{0.0in}\vspace{-1.0in}\\
%\includegraphics[width=1.0\textwidth]{autocor_plq.pdf}
%\rule{0.0in}{0.0in}\vspace{-1.5in}\\
%\caption{
%\label{fig:autocor_plq}
%Autocorrelation in simulation time of the plaquette for the three ensembles.
%Errors on the autocorrelation were estimated by dividing the time series into five
%subsets and averaging the autocorrelations from each subset.
%The vertical arrows indicate the time separation between stored lattices,
%used in computing the potential, spectrum and other quantities.
%}
%\end{figure}
%
%\begin{figure}[t]
%\hspace{-0.1in}
%\rule{0.0in}{0.0in}\vspace{-1.0in}\\
%\includegraphics[width=1.0\textwidth]{autocor_pbp.pdf}
%\rule{0.0in}{0.0in}\vspace{-1.5in}\\
%\caption{
%Autocorrelation in simulation time of the strange quark $\pbp$ in
%the three ensembles.  Here $\pbp$ was estimated using a single
%Gaussian random source vector.
%\label{fig:autocor_pbp}
%}
%\end{figure}

Estimating statistical errors on any physical quantity requires taking into
account the fact that successive sample configurations are not completely
statistically independent, and 
%RS statistical error bars are generally larger
%RS than error estimates made ignoring these autocorrelations.
calculations of statistical errors that ignore these autocorrelations
are generally underestimates of the true errors. 
The amount of autocorrelation depends strongly on the quantity under 
%RS consideration, so here we present autocorrelations for a few simple but
consideration, so we present autocorrelations for a few simple but
relevant quantities.  To parameterize autocorrelations we use the
dimensionless coefficient
\BNE\label{defineauto} C_{\Delta t} = \frac{\LL x_i x_{i+\Delta t} \RR - \LL x_i \RR^2}
                {\LL x_i^2 \RR - \LL x_i \RR^2}
\ENE
where $x_i$ is the measurement at simulation time $i$ and $\Delta t$ is the
time separation of the two measurements.  As discussed above, in these
simulations successive lattices were saved at time separations $\Delta t = 5$
for the $a=0.15$ and $0.12$ fm ensembles, and $\Delta t=6$ for the $0.09$ fm ensemble.
However, measurements of the plaquette and $\pbp$ were made every trajectory.
Note that determination of these autocorrelation coefficients is numerically
difficult, even on time series of order 1000 lattices.
This is partly because of the practical necessity of using the average ($\LL x_i \RR$)
from our simulation, rather than the true average.
(Note, however, that for the topological charge we know that the true average
is zero.)
Estimation of errors
on these coefficients is also noisy.  Here we have estimated the errors from the
variance of autocorrelations measured on five separate segments of the time series,
but for the central value quote the result from the full time series.

%RS These autocorrelations can be handled either by blocking the data (averaging over
The autocorrelations can be taken into account either by blocking the data (averaging over
intervals of time) and then computing the average of the blocked values, or
by multiplying the error estimate ignoring autocorrelations by the factor
\BNE\label{errorstretch} \sqrt{ 1 + 2\sum_t C_t } \ \ ,\ENE
with the sum suitably truncated.
For complicated functions of observables a jackknife analysis
can be used, and Eqs. (\ref{defineauto}) and (\ref{errorstretch}) applied to the
sequence of jackknife results.

%RS I think it would be better to have the table of autocorrelation
%RS results at the end of the section, rather than the beginning,
%RS but as you can see I did not quite succeed. It might be better
%RS to combine the two figures into one with the plaquette and $\pbp4
%RS side by side. Obviously, this is all a matter of taste.

%RS\begin{table}[htb]
\begin{table}[t]
\caption{
\label{tab:autocor}
Autocorrelation $C_{\Delta t}$ of various quantities between successive lattices in the
%RS ensembles.  
%ensembles, or in the case of the plaquette and $\pbp$ between successive
%molecular dynamics trajectories.
%DT changed it back -- the plaquette and PBP autocorrelations are at T=5 and 6
ensembles.
%RS These lattices are separated by five time units for $a=0.15$ and $0.12$
Lattices are separated by five time units for $a=0.15$ and $0.12$
fm, and by six time units for $a=0.09$ fm.
As discussed in the text, the autocorrelations for $\pbp$ are between estimates
made with one random source.
Autocorrelations for the correlators $\LL \pi(0) \pi(D) \RR$ and $\LL \rho(0) \rho(D) \RR$
are given at a spatial distance $D$ which is the minimum distance used in a
typical fit for the mass.   For the pion correlator these distances are
$D=15$, $20$ and $30$ respectively, and for the $\rho$ correlator they are
$D=6$, $7$ and $10$ respectively.
For the pion and rho mass and $f_\pi$ the autocorrelations are from single
elimination jackknife samples.
}
% a=0.09 is [bc] subensembles, since a has different traj. lengths
\begin{tabular}{|c|c|c|c|}
\hline
operator			& $0.15$ fm	& $0.12$ fm	& $0.09$ fm	\\
\hline
$\Box$				& 0.311(25)	& 0.300(10)	& 0.359(14)	\\
$\pbp_{\mathrm{light}}$		& 0.135(24)	& 0.151(27)	& 0.192(34)	\\
$\pbp_{\mathrm{strange}}$	& 0.234(38)	& 0.265(27)	& 0.259(19)	\\
$\LL \pi(0) \pi(D) \RR$		& 0.034(40)	& 0.084(46)	& 0.177(21)	\\
$\LL \rho(0) \rho(D) \RR$	& 0.055(24)	& 0.074(24)	& 0.061(18)	\\
$m_\pi$				& 0.008(14)	& 0.182(35)	& 0.249(51)	\\
$f_\pi$				& 0.123(21)	& 0.150(23)	& 0.184(45)	\\
$m_\rho$			& 0.036(38)	& 0.045(09)	& 0.002(24)	\\
$Q_{\mathrm{topo}}$		& na		& 0.500(25)	& 0.754(36)	\\
\hline
\end{tabular}\end{table}

We begin with the plaquette and strange quark $\pbp$, simple observables
which were measured at each trajectory.
The first two panels of Figure~\ref{fig:autocor_all} show the autocorrelation
of these quantities as a function of separation in simulation time.
Here $\pbp$ is estimated using a single random source vector.  Thus part
of its variance comes from the random source, and part from the variation of
the lattice.  For this reason, its autocorrelation does not approach one
at small time.  We show the strange quark $\pbp$ 
because it generally shows larger autocorrelations
than the light quark $\pbp$.   Also, relevant to future ensembles at other
light quark masses, it will be useful to compare autocorrelations using
$\pbp$ at a fixed physical quark mass.  These two simple quantities provide
a good illustration of how autocorrelations differ among various quantities.

Table \ref{tab:autocor}
shows these quantities at the time separation of
the stored lattices, and a selection of autocorrelations of more physically
relevant quantities.   In particular, it contains autocorrelations of
the pion and rho correlators ($\LL \pi(0) \pi(D) \RR$ and $\LL \rho(0) \rho(D) \RR$)
at a distance $D$
equal to the minimum distance that might be used in a mass fit, and would
be one of the important contributors to the mass.   This table also
contains autocorrelations of single elimination jackknife measurements
of the pion mass, the pion decay constant (amplitude of a pion correlator),
and the rho meson mass.

%The topological charge is generally expected to have a long autocorrelation
%time.  In fact, in the continuum limit tunnelings would be expected to
%be completely suppressed in a simulation algorithm where the configurations
%evolve continuously.  The right panel in Fig.~\ref{fig:autocor_all} shows
%the autocorrelation of the topological charge in the $a=0.12$ and $0.09$ fm
%ensembles.  As expected, the autocorrelation time is larger for this quantity
%than for the others, and is much larger on the finer ensemble.
The topological charge is generally expected to have a long autocorrelation time. In
fact, in the continuum limit tunnelings would be expected to be completely suppressed
in
a
simulation algorithm where the configurations evolve continuously.  Such a simulation
would still
give correct results in infinite volume, but would have power-law finite volume effects
\cite{Leutwyler:1992yt}.
The right panel in Fig. 1 shows the autocorrelation of the topological charge in the
$a = 0.12$ and $0.09$ fm ensembles.
As expected, the autocorrelation time is larger for this quantity than for the others,
and is
much larger on the finer ensemble.  The long autocorrelation time means that it will be
important to check the size of finite volume effects; such runs are planned.

In the sections below, the static quark potential was computed using
block sizes of 50 time units for $a=0.15$ and $0.12$ fm, and 60 time
units for $a=0.09$ fm.  For the pseudoscalar meson plot, block sizes of
20 and 24 time units were used.   Autocorrelations for the rho and nucleon
masses are small, and were neglected here.

\section{The static quark potential}

\begin{figure}[t]
\hspace{-0.1in}
\rule{0.0in}{0.0in}\vspace{-1.0in}\\
%.\vspace{-1.0in}\\
\includegraphics[width=1.0\textwidth]{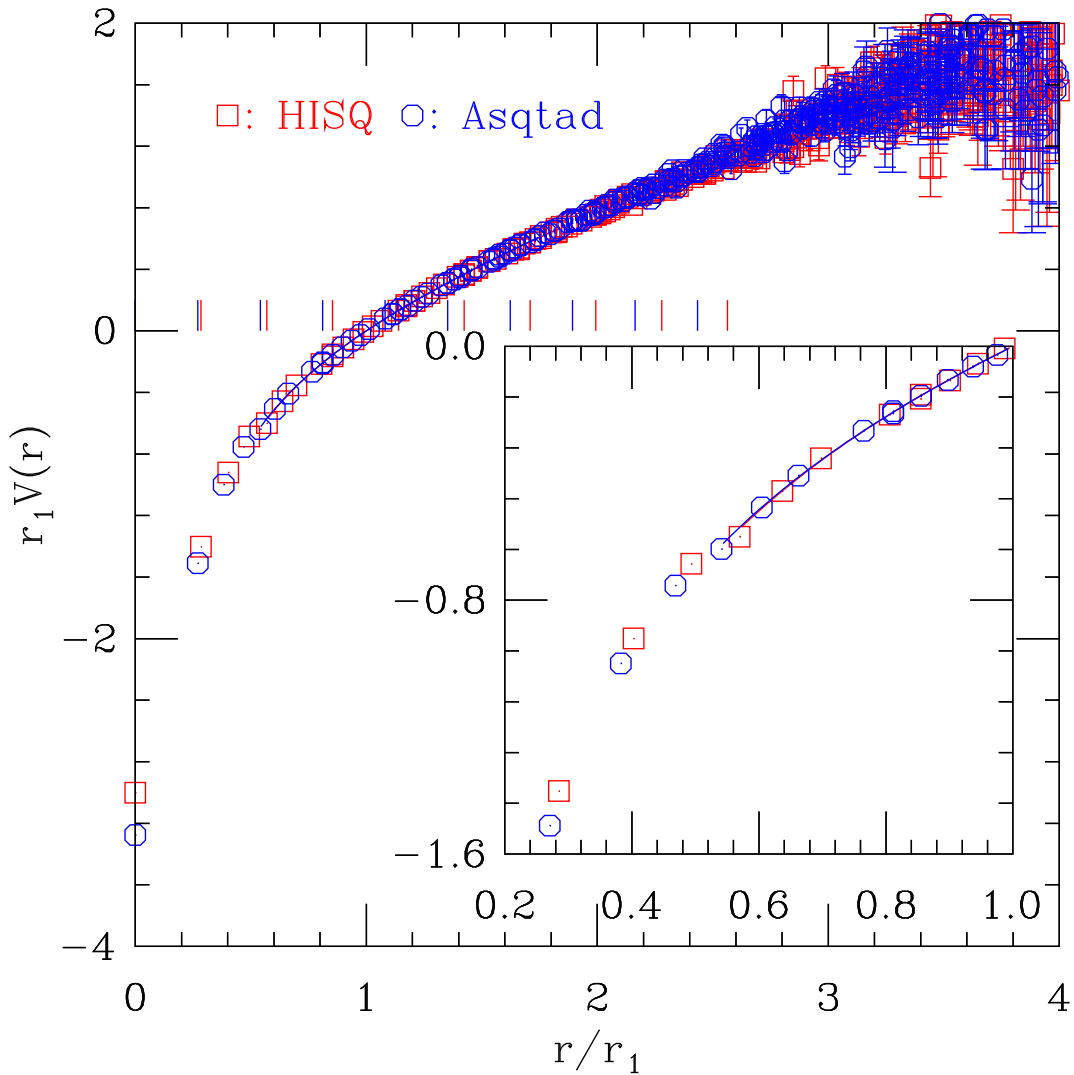}
\rule{0.0in}{0.0in}\vspace{-1.5in}\\
%.\vspace{-1.5in}\\
\caption{
The static quark potential with the HISQ and the asqtad
actions.  The HISQ results are from the $a\approx 0.09$ fm run, and
the asqtad results are from a lattice with almost the same lattice
spacing and light-quark mass about 0.2 times the correct strange
quark mass ($am_l = 0.00465$).  In order to match the potentials, the plot is in
units of $r_1$, while rulers in units of the lattice spacing are shown
at $r_1 V(r) = 0$.  A constant has been added to each potential
so that $V(r_1)=0$. The solid lines (essentially superimposed) show
the fit from Eq.~(\protect\ref{eqn:potform}) for the two runs,
(evaluated with $\lambda$ set to zero).
The inset magnifies a part of this plot at short distance to show
the lattice artifacts discussed in the text.
\label{fig:pot_both}
}
\end{figure}

Although it is not a physical observable, 
the potential between two infinitely heavy test quarks is well defined on
the lattice and can be computed with high precision and comparatively little
effort.  Therefore it has become conventional in lattice simulations to
use a length scale based on the static quark potential to relate
lattice simulations with different couplings, and to translate the
dimensionless results of lattice simulations into physical units.
We generally use $r_1$ defined by $r_1^2 F(r_1) = -1$, where $F(r)$ is the
force $-\PAR{V(r)}{r}$.  The scale $r_0$ defined by $r_0^2 F(r_0) = -1.65$ is
also commonly used.  (The idea behind scales of this form~\cite{SOMMER} is that
they locate the transition region between the Coulomb potential at short
distances, $r^2F(r) = -\frac{4}{3}\alpha$, and the linear potential at
long distances, $r^2 F(r) = -\sigma r^2$.)

In order to determine $r_1$, we measure the static potential at
discrete distances $r^2/a^2 = n_x^2 + n_y^2 + n_z^2$, and, for a range of
$r$ approximately centered at $r_1$, we fit it to the functional form\cite{UKQCD_POTFORM}
\BNE\label{eqn:potform} V(R) = C + \frac{B}{R} + \sigma R +
 \lambda \LP \left. \frac{1}{R} \right|_{lat} - \frac{1}{R} \RP\ \ \ \ . \ENE
%Here $C$ is part of the quarks' self energy, $\sigma$ is the string tension
%and $B$ is $\frac{-3}{4}\alpha_s$ for a potential definition of $\alpha_s$.
Here $\left. \frac{1}{R} \right|_{lat}$ is the 
the lattice Coulomb potential,
$\left. \frac{1}{R} \right|_{lat} = 4\pi \int \frac{d^3p}{(2\pi)^3}
D^{(0)}_{00}(p) e^{ipR}$, with $D^{(0)}_{00}(p)$ the free lattice gluon
propagator calculated with the Symanzik improved
gauge action, and $1/R$ is the continuum Coulomb potential.

Figure~\ref{fig:pot_both} shows the static quark potential at $a\approx 0.09$ fm
for the HISQ ensemble and a corresponding asqtad ensemble.
Overall, the two potentials are very similar.
For reference, the value of $r_1/a$ for this HISQ ensemble in Table~\ref{tab:hisqruns}
came from a fit to the range $\sqrt{5} \le r/a \le 6$, or $0.63 < r/r_1 < 1.70$.
The inset in Fig.~\ref{fig:pot_both} makes visible some of the lattice
artifacts at short distance.   In particular, the HISQ point at $r/r_1=0.57$
and the asqtad point at $0.53$ correspond to separation $(2,0,0)$ along a lattice
axis, and are visibly displaced below the trend.
Note that artifacts of this kind are not decreased with the HISQ action, and
we do not expect them to be decreased. 
In fact, in the continuum limit
we expect them to be described by Eq.~(\ref{eqn:potform}) with $\lambda = B$.
(The fit to this potential has $B=-0.441(6)$ and $\lambda=-0.52(11)$.)
Artifacts like this, at fixed number of lattice spacings, simply move to
$r=0$ in the continuum limit. Also note that these artifacts diminish quickly with
increasing $r$.  For example, the HISQ point at $r/r_1=0.95$ is really two
points, with $\vec r/a = (3,0,0)$ and $(2,2,1)$, and the difference between the
two potential values is invisible.
We expect that these short distance lattice artifacts in the static quark potential
are mostly controlled by the gauge actions, which differ only in the fermion
contributions to the one loop corrections.

We do expect scaling violations proportional to $a^4$ and to $a^2 \alpha^2$
at physical distances for both actions, and these would be visible in quantities like $r_0/r_1$
or $r_1 \sqrt{\sigma}$.  However, it is not possible to make a definitive
comparison of scaling violations in these quantities between the two actions
yet, since the addition of the dynamical charm quark to the HISQ simulations
could also have small effects on these quantities.

For reference, Table~\ref{tab:potfits} shows the parameters of the
fits in Fig.~\ref{fig:pot_both}, defined in Eq.~(\ref{eqn:potform}).
Note that in this figure the fitting range used for the HISQ run is the
same as used for the asqtad ensemble, and so differs
from that used in finding the value of $r_1/a$ in Table~\ref{tab:hisqruns}.
Also note that the quantity $\sigma r_1^2$ parameterizes the potential
in the range around $r_1$, and should not be used as a measurement of the
long distance string tension.   Finally, note that since the dimensionful
parameters are expressed in units of $r_1$, which is found from the same
fit, one relation between $B$ and $\sigma$ is automatically enforced.
In Fig.~\ref{fig:pot_both} this constraint forces both fits to have
the same slope at $r=r_1$ (since $r_1$ is defined by the slope (force)
at this distance), and a constant was subtracted to make both fits be zero
at this point.

\begin{table}
\caption{
\label{tab:potfits}
Parameters of the potential fits in Fig.~\protect\ref{fig:pot_both}.
As discussed in the text, in this comparison the fit ranges for the HISQ
potential were chosen to match those used for the asqtad potential, and so
these tabulated parameters differ slightly from those used in the rest of
this paper. Note that the lattice mass is regularization dependent  --- in both
of these ensembles the light quark mass
is about one fifth of the correct strange quark mass.
}
\begin{tabular}{|l|l|l|}
\hline
		& asqtad	&	HISQ	\\
\hline
Fit range	& 2.01--6.5	& 2.01--6.5	\\
Time separations & 5--6		& 5--6		\\
$10/g^2$	& 7.085		& 6.30 \\
$am_l/am_s/am_c$ & 0.00465/0.031/na & 0.0074/0.037/0.440 \\
$C a$		& 0.849(3)	& 0.824(3) \\
$B$		& $-$0.432(4)	& $-$0.450(5) \\
$\sigma r_1^2$	& 0.568(6)	& 0.554(6) \\
$r_1/a$		& 3.697(7)	& 3.510(7)	\\
\hline
\end{tabular}
\end{table}

\section{Scaling tests}

\begin{figure}[t]
\hspace{-0.1in}
\rule{0.0in}{0.0in}\vspace{-1.0in}\\
\includegraphics[width=1.0\textwidth]{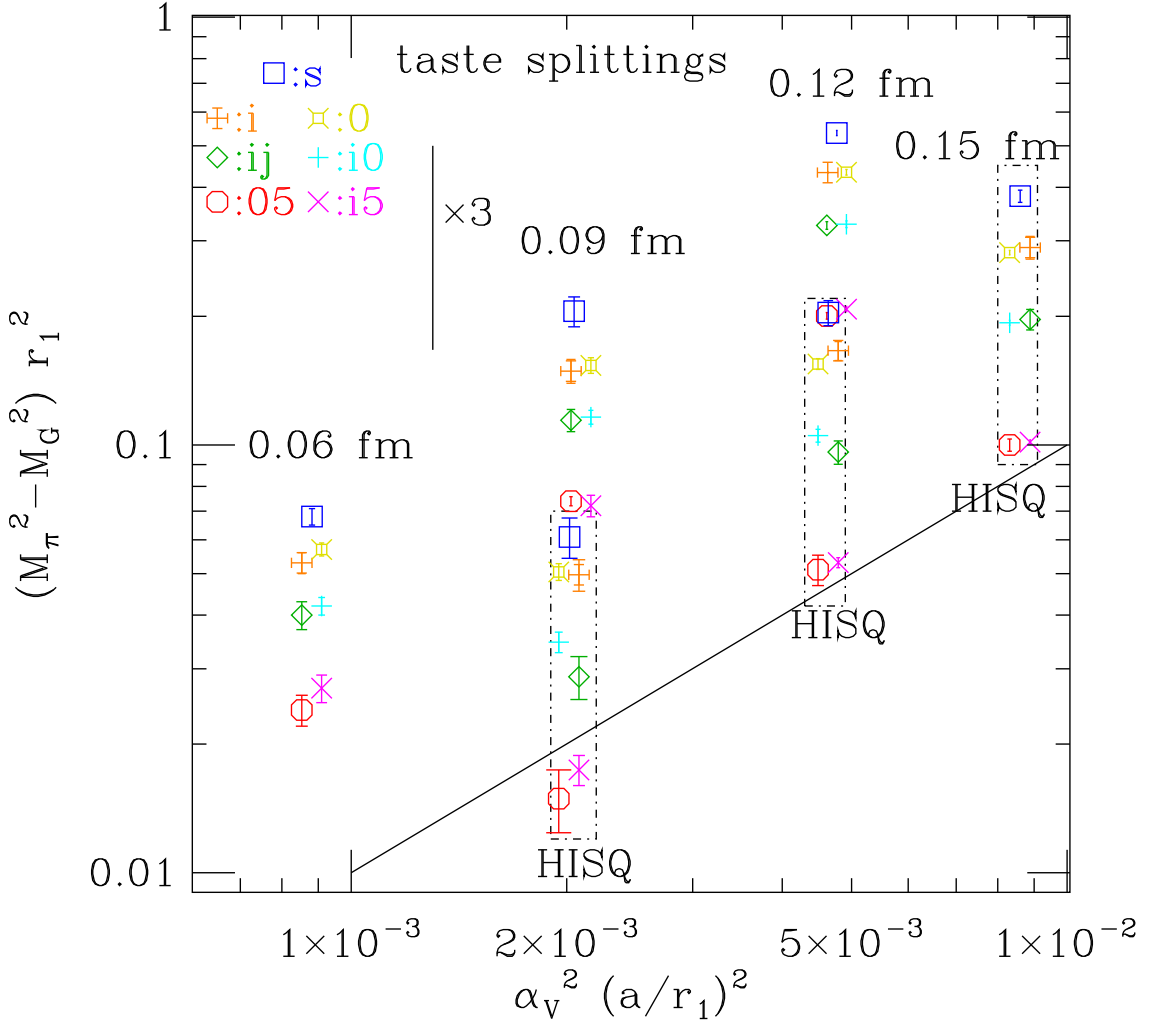}
\rule{0.0in}{0.0in}\vspace{-1.5in}\\
\caption{Taste splittings among the pions.  The asqtad results used
configurations with 2+1 flavors of dynamical quarks, and the HISQ results
2+1+1 flavors. The
quantity plotted is $r_1^2 \LP M_\pi^2 - M_G^2 \RP$, where $M_\pi$
is the mass of the non-Goldstone pion and $M_G$ is the mass of the
Goldstone pion.   
%RS This quantity is known to be almost independent of
%RS the light-quark mass.
%RS Note this sentence below is new and the symbols need to be checked.
Reading from top to bottom, the non-Goldstone pions
are the $\pi_s$ (box), $\pi_0$ (fancy box), $\pi_i$ (fancy plus),
$\pi_{io}$ (plus), $\pi_{ij}$ (diamond), $\pi_{i5}$ (cross)
and $\pi_{05}$ (octagon).
$r_1^2 \LP M_\pi^2 - M_G^2 \RP$ is known to be almost independent of
the light-quark mass.
The vertical bar at the upper left shows the size of a factor of
three, roughly the observed reduction in taste splittings, while
the sloping solid line shows the theoretically expected dependence on lattice
spacing.  Nearly degenerate points have been shifted horizontally to improve
their visibility.
\label{fig:taste_splitting}
}
\end{figure}

Reduction of taste splittings among the pion masses with HISQ valence quarks
was demonstrated
with quenched gauge fields in Refs.~\cite{hpqcd_hisq2003,hpqcd_hisq2004}, and with asqtad
sea quarks in Ref.~\cite{hpqcd_hisqprd}, and there
is little reason to expect it to be different with dynamical HISQ sea
quarks.   However, in view of the importance of this quantity, we show
splittings for all of the different tastes of pions in
Fig.~\ref{fig:taste_splitting}, comparing results with HISQ quarks (both
valence and sea) to earlier results with asqtad quarks.
In this figure, we see that the expected reduction in taste splittings
happens, with roughly a factor of three reduction relative to asqtad
calculations at the same lattice spacing.

\begin{figure}[t]
\hspace{-0.1in}
\rule{0.0in}{0.0in}\vspace{-1.0in}\\
\includegraphics[width=1.0\textwidth]{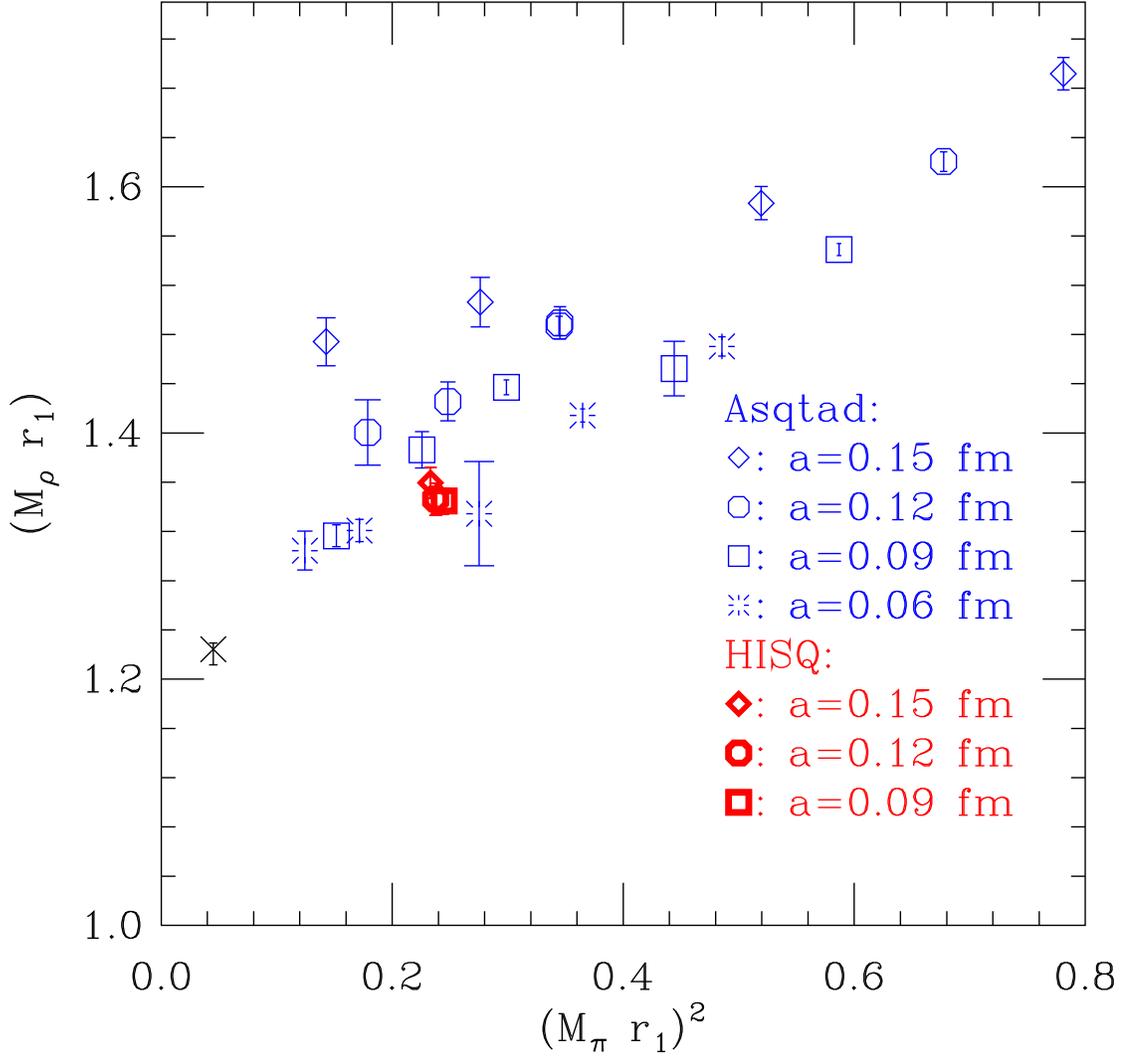}
\rule{0.0in}{0.0in}\vspace{-1.5in}\\
\caption{Vector meson ($\rho$) masses in units of $r_1$.
Here the bold (red) points are the HISQ simulations with $m_l=0.2 m_s$,
and the lighter (blue) points are asqtad results for various light
quark masses.  
%RS The sentence below has been added.
The $a\approx 0.06$~fm asqtad point immediately to the right of the
$a\approx 0.09$~fm HISQ point has been displaced to the right to make it 
visible.  It in fact falls on top of the $a\approx 0.09$~fm HISQ point.
The cross sign at lower left is the physical
$\rho$ mass.  The error on the physical mass point is just the error
on the physical value of $r_1$.
\label{fig:rhomass}
}
\end{figure}

\begin{figure}[t]
\hspace{-0.1in}
\rule{0.0in}{0.0in}\vspace{-1.0in}\\
\includegraphics[width=1.0\textwidth]{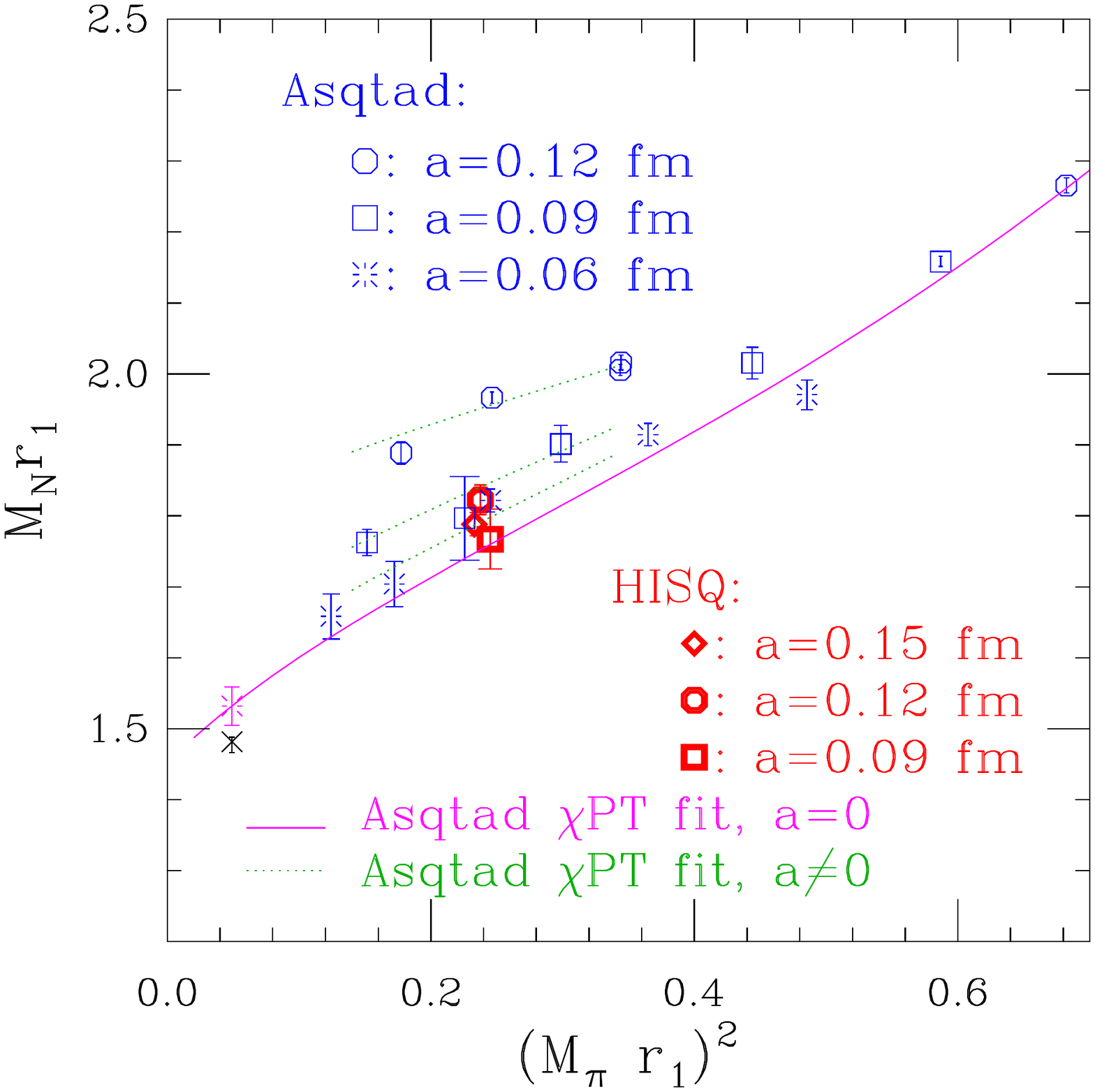}
\rule{0.0in}{0.0in}\vspace{-1.5in}\\
\caption{Nucleon masses in units of $r_1$.
Here the bold (red) points are the HISQ simulations with $m_l=0.2 m_s$,
and the lighter (blue) points are asqtad results for various light
quark masses.  The cross at lower left is the physical
nucleon mass.  The solid magenta line is a continuum extrapolation of a
chiral perturbation theory fit to the asqtad nucleon masses,
while the dotted green lines are from the same fit at finite lattice
spacing~\protect\cite{lat07_baryons}.
\label{fig:nucmass}
}
\end{figure}

The main purpose of this study was to see if the improvements in the
action designed to reduce taste symmetry violations translate into 
decreased lattice spacing dependence in other quantities.  We begin with
the mass of the light-quark vector meson, or $\rho$.  In Fig.~\ref{fig:rhomass}
we show the mass of the $\rho$ meson in units of $r_1$.
Here we have asqtad results for several light-quark masses at
each lattice spacing, but HISQ results for only one light-quark
mass.  The light-quark masses themselves are regularization dependent,
so to plot asqtad and HISQ results on the same footing we use the
Goldstone pion mass in units of $r_1$ for the horizontal axis.
Note that for $m_l=0.2 m_s$, the light-quark mass used in the HISQ
simulations, and for the lattice sizes used here ($\le 2.9$ fm),
the vector meson is stable against decay to two pions.
Results for the nucleon mass are similar, and are shown in Fig.~\ref{fig:nucmass}.
In Figs.~\ref{fig:rhomass} and \ref{fig:nucmass} the HISQ masses show smaller
dependence on the lattice spacing than the asqtad masses, with the same continuum limits within
the statistical errors.  Roughly speaking, the HISQ results are similar to
the asqtad results at the next smaller lattice spacing.

Although we have chosen to present these results as improved scaling
of the $\rho$ and nucleon masses, since we are plotting the dimensionless
quantities $M_\rho r_1$ and $M_N r_1$, they could equally well be described as improved
scaling of $r_1$ when a hadron mass is chosen to be the
length standard.

\begin{figure}[t]
\hspace{-0.1in}
\rule{0.0in}{0.0in}\vspace{-1.0in}\\
\begin{tabular}{ll}
\includegraphics[width=0.5\textwidth]{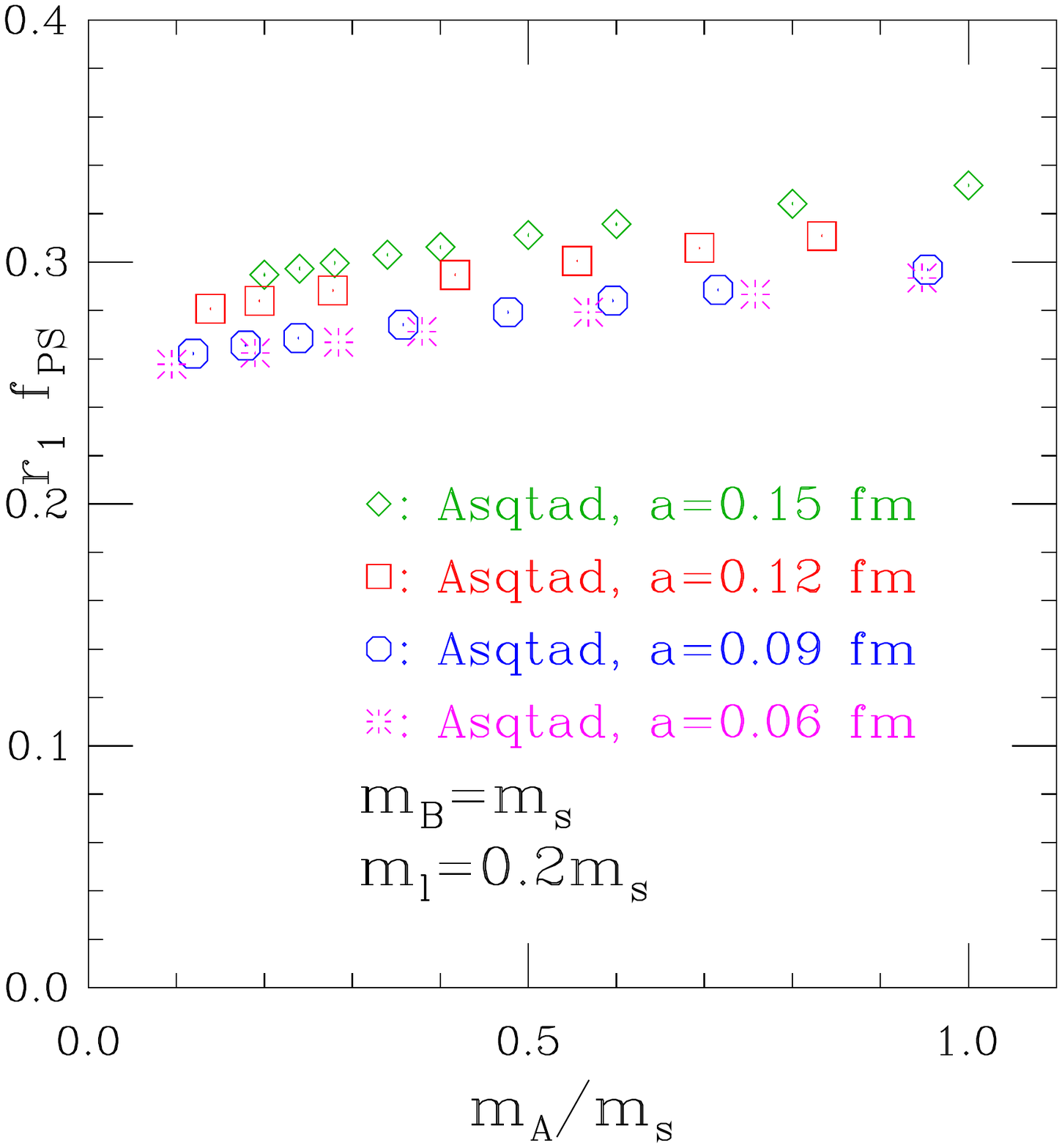} &
\includegraphics[width=0.5\textwidth]{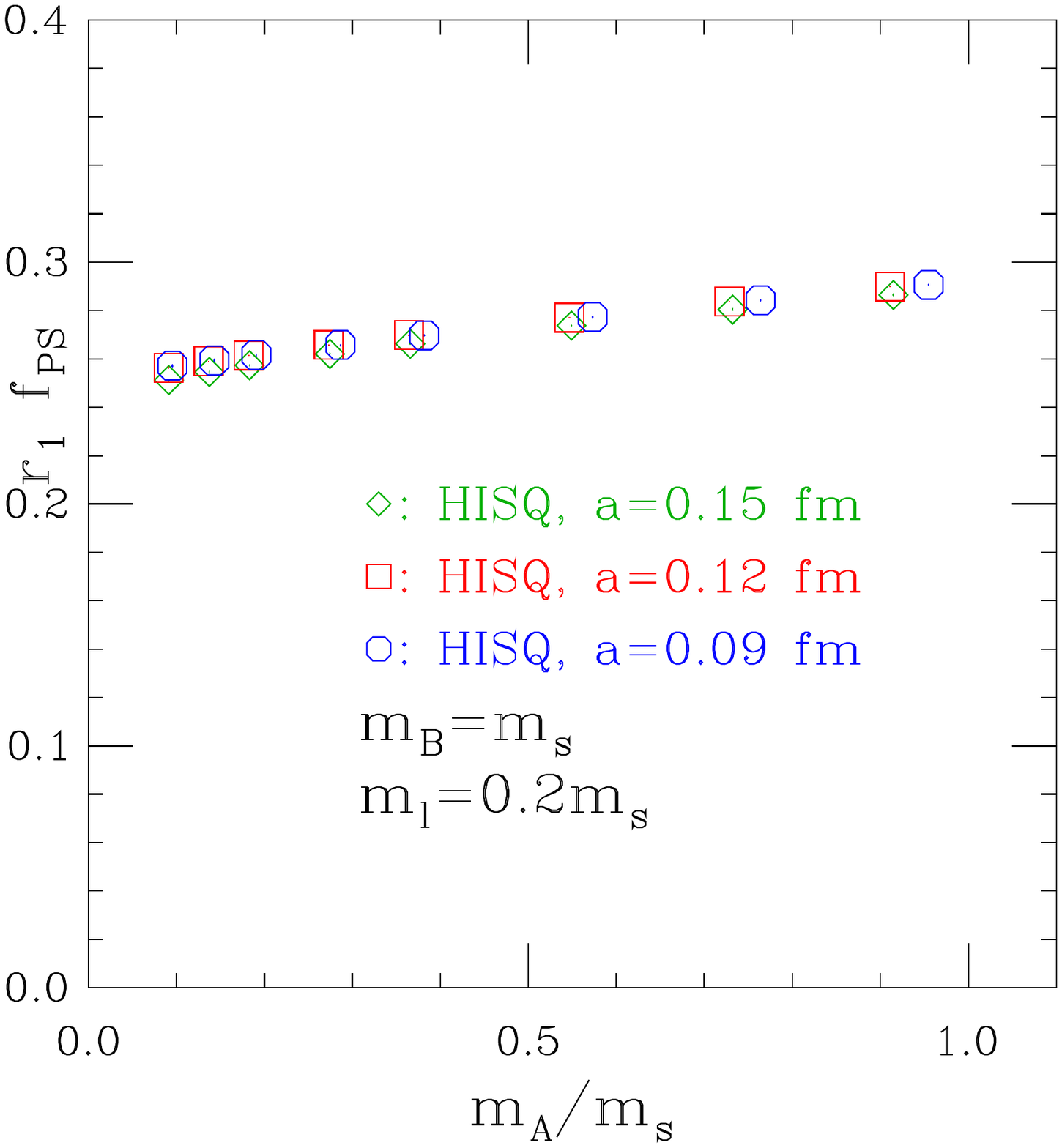} \\
\end{tabular}
\rule{0.0in}{0.0in}\vspace{-1.0in}\\
\caption{Pseudoscalar decay constant.  One valence-quark mass, $m_A$,
is varied while the second is held fixed near the strange-quark
mass.  All ensembles used a light sea quark mass of about
$0.2$ times the strange-quark mass.   The left hand panel
shows asqtad results for four different lattice spacings
and the right hand panel shows HISQ results for three lattice
spacings.
\label{fig:fpi_both}
}
\end{figure}

\begin{figure}[t]
\hspace{-0.1in}
\rule{0.0in}{0.0in}\vspace{-1.0in}\\
\includegraphics[width=1.0\textwidth]{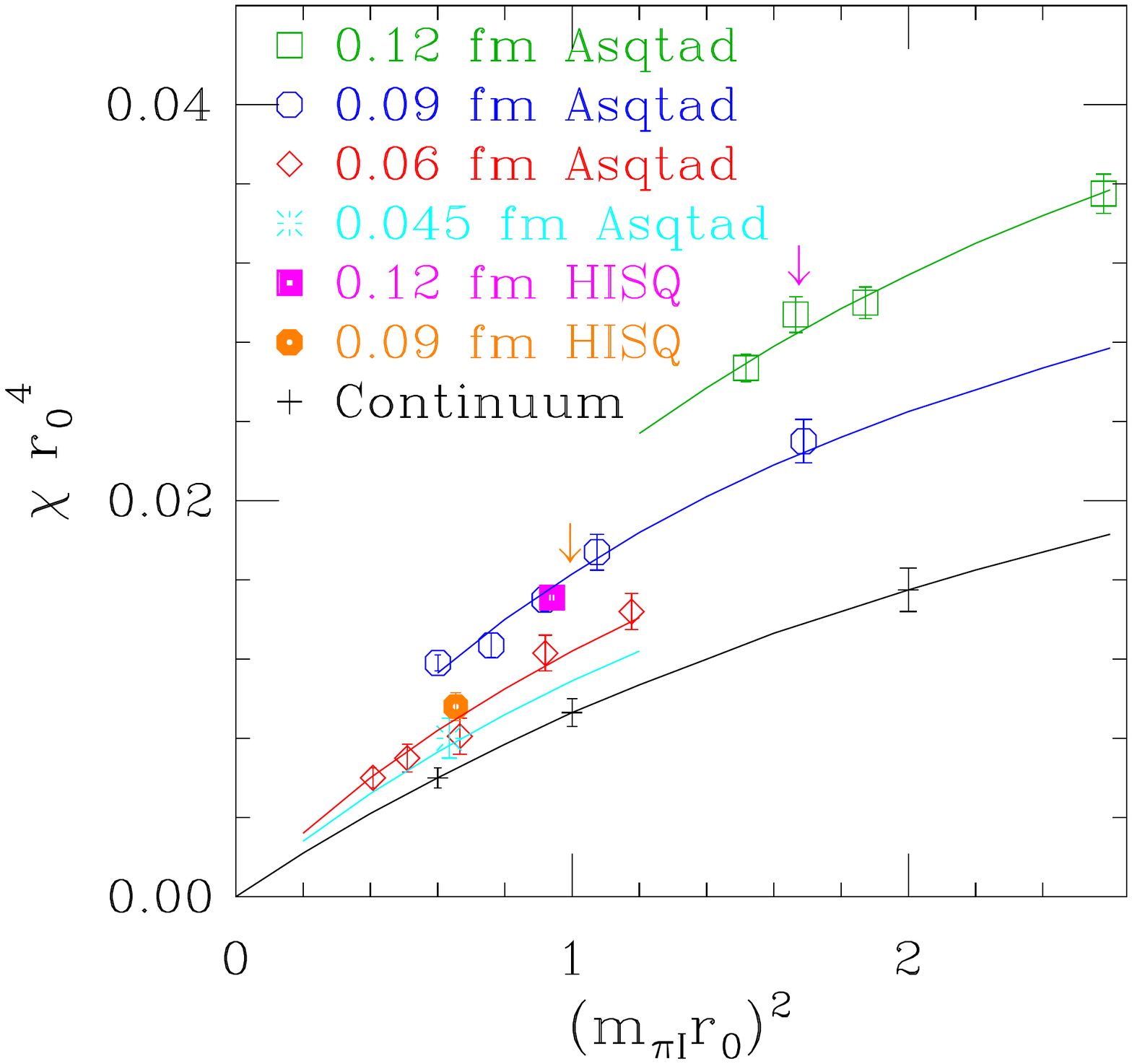}
\rule{0.0in}{0.0in}\vspace{-1.5in}\\
\caption{ %{\bf LINEAR SCALE VERSION}
The topological susceptibility.  Points with the
asqtad action are shown for several lattice spacings and quark
masses, and the
HISQ results for $a \approx 0.12$ fm and $a \approx 0.09$ fm with $m_l = 0.2 m_s$.
For the horizontal axis we use the mass of the taste singlet pion, since
in lowest order chiral perturbation theory the topological susceptibility
is a function of this mass~\protect\cite{topo_form}.
The curves in the
figure come from a chiral perturbation theory fit to the asqtad data.
The asqtad results are updated from Ref.~\protect\cite{topo_lat07} and
are discussed further in Refs.~\protect\cite{milc_rmp,topo_milc10}.
The two arrows indicate the locations of asqtad points with lattice spacing and
quark mass similar to the two HISQ points. (In the case of the $a\approx 0.09$ fm
HISQ point, the quark mass falls between two of the masses of the asqtad points.)
\label{fig:topo_susc}
}
\end{figure}

The pseudoscalar meson decay constants are important for lattice
determinations of CKM matrix elements, and can be computed with high
precision.  In fact, our current best determination of $r_1$ in physical
units comes from matching the asqtad lattice results to the physical
value of $f_\pi$.  These decay constants for light quarks have been
extensively studied using the asqtad ensembles~\cite{milc_asqtad_fpi,milc_rmp}.
The HPQCD collaboration has computed these decay constants in a
mixed action calculation, with HISQ valence quarks on the asqtad
sea quark ensembles, and used them in a determination of the
physical value of $r_1$ ~\cite{hpqcd_hisq_r1}.
Figure~\ref{fig:fpi_both} shows the pseudoscalar decay constant
with one of the valence quarks fixed at approximately the strange
quark mass as a function of the mass of the other valence quark.
(At the physical light-quark mass, this is just $f_K$.)
To facilitate the comparison, we have used the ratio of the
light quark mass to the corrected strange quark mass in the corresponding
ensemble for the horizontal axis.
%DT changed plot axis to m_l/m_s, so all this is unneccessary
%
%Since the quark mass is regularization dependent, the horizontal
%scale for the asqtad results (left panel) has been rescaled by a
%factor $Z_{rel}$ determined from the quark masses that give
%the physical pseudoscalar masses.   For example, at $a \approx 0.12$
%fm, the correct asqtad strange-quark mass is $a m_s = 0.036$, while
%for HISQ it is $a m_s = 0.0509$.   Thus the asqtad quark masses
%were rescaled by $Z_{rel} = 0.0509/0.036$.
%(Note that our normalization for the asqtad quark masses differs
%by a factor of $u_0$ from that used by the HPQCD collaboration:
%$am_q(\mathrm{MILC}) = u_0 a m_q(\mathrm{HPQCD})$.  Since we do not use tadpole improvement
%in the HISQ fermion action, the two groups have the same
%definition of the HISQ quark mass.)
The reduction in lattice artifacts is obvious, and it can also
be seen that the HISQ points lie near the continuum limit of the
asqtad points.   Once again, we remark that since the plotted
quantity is $r_1 f_{PS}$, this could equally well be described as
scaling of $r_1$ or scaling of $f_{PS}$.

The topological susceptibility is a particularly important test
here, since it is computed solely from the gluon configurations that
are generated, without involving HISQ or asqtad valence quarks.
Therefore improvements in the scaling of the topological susceptibility
directly test whether the change of the sea-quark action has the
expected effect on the gluon configurations that are generated.
%RS The following two lines are new. I switched the order of Refs
%RS topo_lat07 and topo_milc10 because the latter is now mentioned first.
Our technique for calculating the topological susceptibility
is set out in detail in Ref.~\cite{topo_milc10}.
%DT added a sentence to emphasize this
Here we just note that this technique is based on measurement of a
density-density correlator, and hence is not limited by long autocorrelation
times for the overall topological charge.
Figure~\ref{fig:topo_susc} shows the topological susceptibility for
most of the asqtad ensembles, and HISQ results for the $a\approx 0.12$
fm  and $a \approx 0.09$ fm ensembles.
The HISQ point with $a\approx 0.12$ fm lies near the asqtad points
with $a\approx 0.09$ fm, and the HISQ point with $a\approx 0.09$ fm is near
the asqtad points with $a\approx 0.06$ fm, demonstrating a decrease in lattice artifacts.
Note that the HISQ points are to the left of the corresponding asqtad points,
which are indicated by arrows in the figure.
This is because the horizontal axis is the mass of the taste singlet
pion (the heaviest pion taste), and the reduction in taste symmetry
breaking moves the points to the left.  It is the movement down relative
to the asqtad points that represents an improvement in the gluon configurations.

%\begin{figure}[t]
%\hspace{-0.1in}
%\includegraphics[width=0.8\textwidth]{topology_dec09_log.pdf}
%\caption{{\bf LOG SCALE VERSION} The topological susceptibility.  Points with the
%asqtad action are shown for several lattice spacings and quark
%masses, and the
%HISQ results for $a \approx 0.12$ fm and $a \approx 0.09$ fm with $m_l = 0.2 m_s$.
%For the horizontal axis we use the mass of the taste singlet pion, since
%in lowest order chiral perturbation theory the topological susceptibility
%is a function of this mass~\protect\cite{topo_form}.
%The asqtad results are updated from Ref.~\protect\cite{topo_lat07} and
%are discussed further in Ref.~\protect\cite{milc_rmp}.
%The two arrows indicate the asqtad points with lattice spacing and
%quark mass similar to the two HISQ points.
%\label{fig:topo_susc}
%}
%\end{figure}

\begin{table}[t]
\caption{
\label{tab:fss_vs_r1}
Lattice spacings in fm from $r_1=0.3117$ fm, 
$f_{ss}$ with asqtad valence quarks,
and $f_{ss}$ with HISQ valence quarks.
The first five columns identify the ensemble by the sea-quark action, the gauge
coupling $10/g^2$, and the sea-quark masses.
The horizontal line separates ensembles with asqtad sea quarks (above) from those
with HISQ sea quarks (below).
The values for HISQ valence quarks on asqtad sea ensembles
are taken from Ref.~\protect\cite{hpqcd_hisq_r1}.
The errors on $a(r_1)$ are statistical only --- they do not include the 
errors in $r_1=0.3117(6)(\null_{-31}^{+12})$ fm. 
Similarly, the errors on $a(f_{ss}-\mathrm{asqtad})$ and $a(f_{ss}-\mathrm{HISQ})$ for the HISQ ensembles
do not include any errors in the physical value of $f_{ss}$.
The numbers following the $f_{ss}$ lattice spacings are the
value of the valence strange quark mass $a m_s$ at which the desired ratio is obtained.
We use the values $f_{ss} = 181.5$ MeV and $f_{ss}/M_{ss} = 0.2647$
from Ref.~\protect\cite{hpqcd_hisq_r1}.
}
% fixed error in asqtad 7.09 line, 6/1/10 DT
%\begin{tabular}{|llllc|lcl|}
%\hline
%Action & $10/g^2$ & $am_l$ & $am_s$ & $am_c$ 	& $a(r_1)$	& $a(f_{ss}-\mathrm{asqtad})[am_s]$	&
%$a(f_{ss}-\mathrm{HISQ})[am_s]$\\
%\hline
%%b6572m0097m04845 & 0.1453(9)	& 0.1770(4)	& 0.1583(13) \\
%% quadratic solution doesn't work for this one
%asqtad & 6.76 & 0.01   & 0.05  & -- & 0.1178(2)	& 0.1373(2)[0.0467]	& 0.1264(11)[0.0553] \\
%asqtad & 7.09 & 0.0062 & 0.031 & -- & 0.0845(1)	& 0.0905(3)[0.0286]	& 0.0878(7)[0.0362] \\
%asqtad & 7.46 & 0.0036 & 0.018 & -- & 0.0588(2)	& 0.0607(1)[0.0187]	& 0.0601(5)[0.0233] \\
%asqtad & 7.81 & 0.0028 & 0.014 & -- & 0.0436(2)	& 0.0444(1)[0.0133]	& 0.0443(4)[0.0163] \\
%\hline
%HISQ & 5.80 & 0.013  & 0.065  & 0.838 & 0.1527(7)	& na	& 0.1558(3)[0.0720] \\
%HISQ & 6.00 & 0.0102 & 0.0509 & 0.635 & 0.1211(2)	& na	& 0.1244(2)[0.0549] \\
%HISQ & 6.30 & 0.0074 & 0.037  & 0.440 & 0.0884(2)	& na	& 0.0900(1)[0.0374] \\
%%b672m0048m024m286	& 0.0588(4?)	& na	& 0.0602(1) \\
%	\hline
%	\end{tabular} \\
\begin{tabular}{|llllc|l|cc|cc|}
\hline
Action & $10/g^2$ & $am_l$ & $am_s$ & $am_c$ 	& $a(r_1)$	& $a(f_{ss}-\mathrm{asqtad})$ & $am_s$	&
$a(f_{ss}-\mathrm{HISQ})$ & $am_s$\\
\hline
%b6572m0097m04845 & 0.1453(9)	& 0.1770(4)	& 0.1583(13) \\
% quadratic solution doesn't work for this one
asqtad & 6.76 & 0.01   & 0.05  & -- & 0.1178(2)	& 0.1373(2) & 0.0467	& 0.1264(11) & 0.0553 \\
asqtad & 7.09 & 0.0062 & 0.031 & -- & 0.0845(1)	& 0.0905(3) & 0.0286	& 0.0878(7) & 0.0362 \\
asqtad & 7.46 & 0.0036 & 0.018 & -- & 0.0588(2)	& 0.0607(1) & 0.0187	& 0.0601(5) & 0.0233 \\
asqtad & 7.81 & 0.0028 & 0.014 & -- & 0.0436(2)	& 0.0444(1) & 0.0133	& 0.0443(4) & 0.0163 \\
\hline
HISQ & 5.80 & 0.013  & 0.065  & 0.838 & 0.1527(7)	& na	& na	& 0.1558(3) & 0.0720 \\
HISQ & 6.00 & 0.0102 & 0.0509 & 0.635 & 0.1211(2)	& na	& na	& 0.1244(2) & 0.0549 \\
HISQ & 6.30 & 0.0074 & 0.037  & 0.440 & 0.0884(2)	& na	& na	& 0.0900(1) & 0.0374 \\
%b672m0048m024m286	& 0.0588(4?)	& na	& 0.0602(1) \\
	\hline
	\end{tabular} \\
\end{table}

\section{Using $\bf f_{ss}$ to set the scale}

%In Figs.~\ref{fig:rhomass}, \ref{fig:nucmass} and \ref{fig:fpi_both} it can be seen
In Figs.~\ref{fig:rhomass}--\ref{fig:fpi_both} it can be seen
that $r_1 f_K$ and the hadron masses in units of $r_1$
%are all approaching their continuum limits from above as $a\rightarrow 0$.
all increase as the lattice becomes coarser.
This common dependence on lattice
spacing could be absorbed into a lattice spacing dependence of $r_1$.
Put more simply, we could use one of these quantities to set the
lattice spacing.  Such a procedure has been introduced and studied
by the HPQCD collaboration in Ref.~\cite{hpqcd_hisq_r1}.  In particular,
%one can take the decay constant of a fictitious ``unmixed $\bar s s$'' pseudoscalar
%meson, which is an isospin non-singlet meson with both quarks having
%mass $m_s$, and declare that this quantity has no lattice spacing or
%sea-quark mass dependence.  Like $r_1$, this is not a quantity that can be directly
%determined from experiment, and so, like $r_1$, its physical value
%is eventually determined by matching to some precisely known quantity
%such as $f_\pi$ or mass splittings of heavy quark mesons.  However,
%we can begin with good (order 1\% error) estimates of this quantity
%from chiral perturbation theory.  This quantity, which we call $f_{ss}$,
%can be determined to high precision in the lattice simulations.
they use the decay constant of a fictitious ``unmixed $\bar s s$'' pseudoscalar
meson, which is an isospin non-singlet meson with both valence quarks having
mass $m_s$, to set the scale. We call this decay constant $f_{ss}$.  Like $r_1$, $f_{ss}$
is not a quantity that can be directly
determined from experiment, and so, like $r_1$, its physical value
is eventually determined by matching to some precisely known quantity
such as $f_\pi$ or mass splittings of heavy quark mesons. In practice, the HPQCD collaboration
determines $f_{ss}$ and the corresponding meson mass $M_{ss}$ 
using a next-to-leading-order chiral fit (augmented with discretization 
corrections) to their lattice data, and inputs of the experimental values for 
$f_\pi$, $f_K$, $M_\pi$ and $M_K$. In fact, lowest order chiral perturbation theory and these
experimental values alone (without lattice data) gets within $\sim 1\%$ of the HPQCD
results \cite{hpqcd_hisq_r1}.
We prefer not to input the experimental value $f_K$ in such determinations, since
we take $f_K$ as an output of our lattice calculations that gives a result 
for $V_{us}$ \cite{milc_asqtad_fpi,milc_rmp}.
Indeed, $f_\pi$, $M_\pi$ and $M_K$ alone are adequate for determining the physical scale
and the quark masses $m_l$ and $m_s$, and hence all light-quark quantities. 

An advantage of using $f_{ss}$ to set the scale on a given lattice ensemble is that it can
be determined to high accuracy in the simulations.
However, it has the disadvantage that it depends on the choice of
valence quarks, so the lattice spacing assigned to a particular ensemble will
depend slightly on whether it is determined with asqtad quarks, HISQ
quarks, or some other formalism.

\begin{figure}[t]
\hspace{-0.1in}
\rule{0.0in}{0.0in}\vspace{-1.0in}\\
\includegraphics[width=1.0\textwidth]{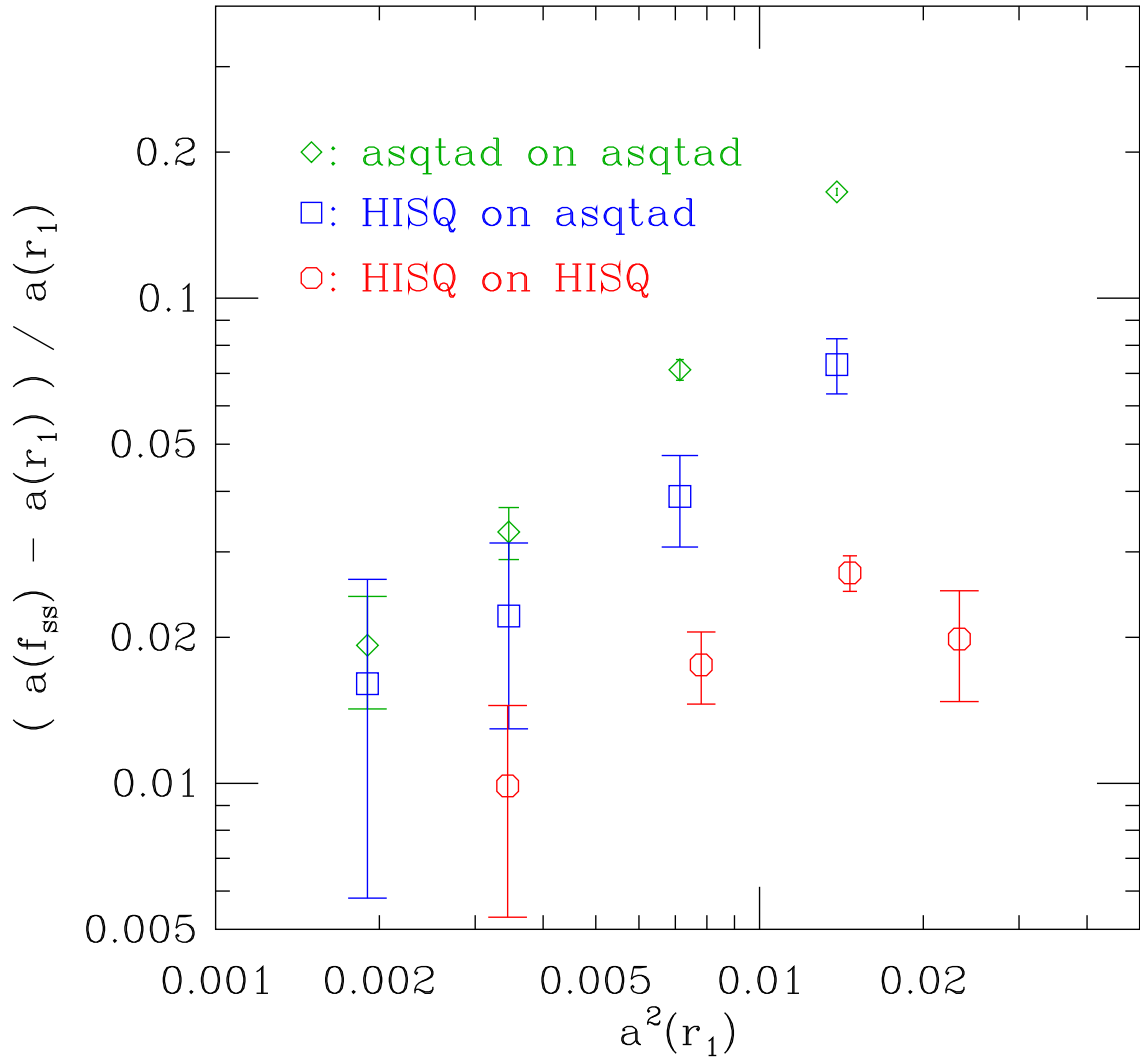}
\rule{0.0in}{0.0in}\vspace{-1.5in}\\
\caption{
Differences in determinations of the length scale using different
standards.  In the legend, the symbol types are labelled as
``valence on sea''.
%The ``FP4S'' points (bursts) use a pseudoscalar
%meson with both sea quarks at 0.4 times the strange quark mass, while
%all the other points use valence quarks at the strange quark mass.
The ``HISQ on asqtad'' points are taken from Ref.~\protect\cite{hpqcd_hisq_r1}.
\label{fig:find_a_2}
}
\end{figure}

\begin{figure}[t]
\hspace{-0.1in}
\rule{0.0in}{0.0in}\vspace{-1.0in}\\
\includegraphics[width=1.0\textwidth]{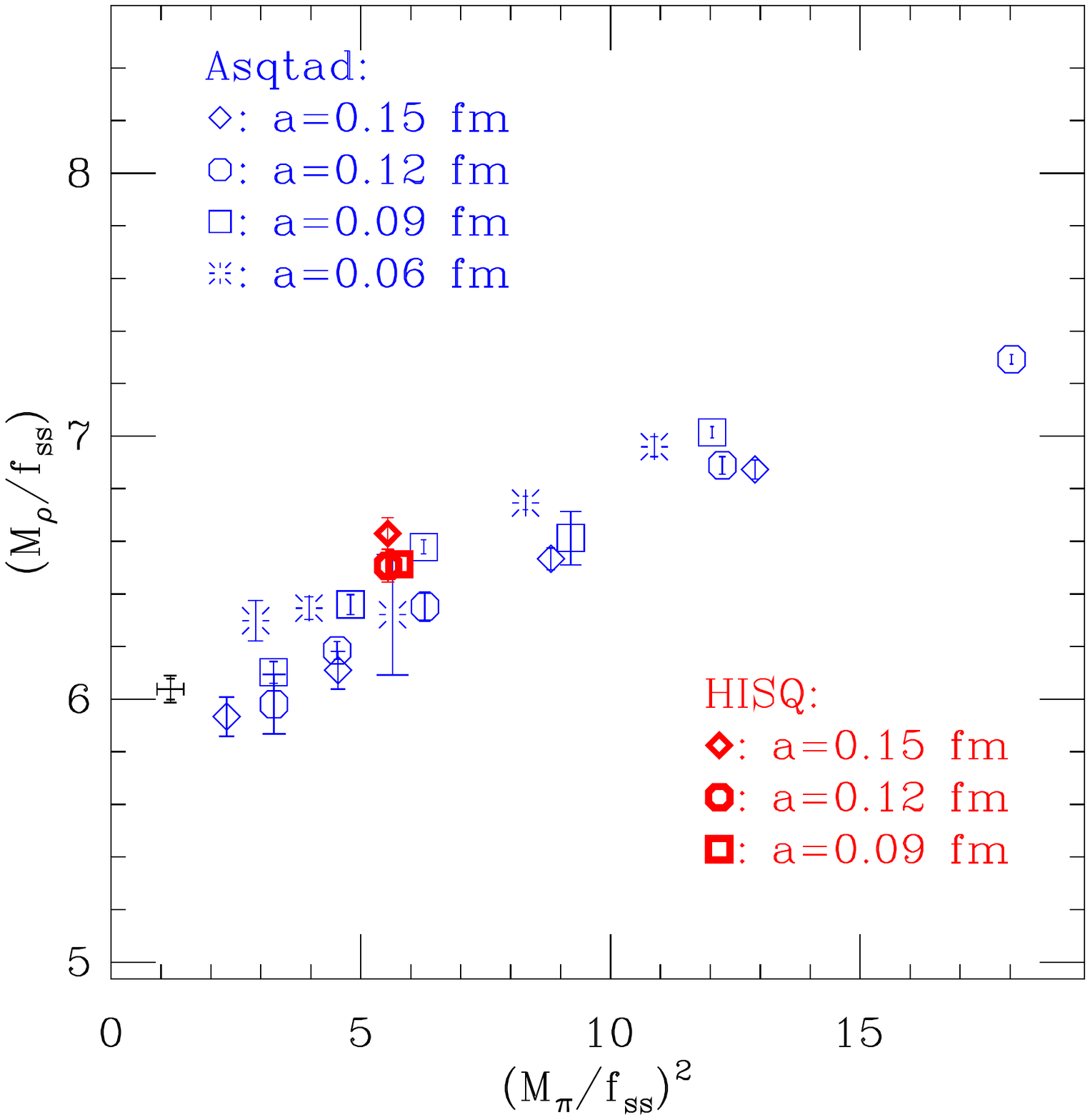}
\rule{0.0in}{0.0in}\vspace{-1.5in}\\
\caption{Vector meson ($\rho$) masses in units of $f_{ss}$.
The data and the meaning of the symbols are
the same as in Fig.~\protect\ref{fig:rhomass}.
The vertical and horizontal scales in the figure
correspond to the same ranges as in Fig.~\protect\ref{fig:rhomass}.
\label{fig:rhomass_fss}
}
\end{figure}

Table~\ref{tab:fss_vs_r1} shows the lattice spacings of the three
HISQ ensembles used in this paper, and some comparable asqtad
ensembles, using $r_1$ and $f_{ss}$ as the length standards.
For the asqtad ensembles, we show the effect of using either
asqtad or HISQ valence quarks to determine the lattice spacing.
Note that, as expected, the differences among the scale determinations
decrease as the lattice spacing decreases.
The value of $f_{ss}$ and the corresponding strange quark mass $a m_s$
were determined by fitting a quadratic polynomial through masses and decay
constants at valence masses equal to 1.0, 0.8 and 0.6 times the sea
strange quark mass.
Table~\ref{tab:fss_vs_r1} also shows the value of the strange quark mass $a m_s$ given by this
interpolation or extrapolation.
Figure \ref{fig:find_a_2} shows the differences in length scale (relative to
the determination from $r_1$) as a function of lattice spacing.  In this figure
it can be seen that these differences are vanishing in the expected way as
the lattice spacing decreases.

In Fig.~\ref{fig:rhomass_fss} we show the rho mass data from Fig.~\ref{fig:rhomass}
replotted using $f_{ss}$ to set the scale.  Replotting the nucleon masses
in Fig.~\ref{fig:nucmass} would produce similar results.  (Of course, one
could then make a plot showing the dependence of $r_1 f_{ss}$ on sea-quark mass
and lattice spacing.)

% f_ss to set the lattice spacing
% note that f_pi and hadron masses are all increasing with lattice
% spacing from r_1.   We could use one of them as length scale, and
% say r_1 is dependent on lattice spacing.
% define it.  reference HPQCD
% rho scaling plot  (note f_pi plot necessarily good, since
% we defined it to be perfect at one point.)
% table of relative lattice spacings.
% ref is hpqcd_hisq_r1

\section{Conclusions}

Using simulations with a fixed and unphysically large light-quark
mass, we see that dimensionless ratios of several hadronic quantities
show smaller dependence on lattice spacing with the HISQ action than
with the asqtad action.  
Roughly, for the quantities that we checked,
HISQ simulations at lattice spacing $a$ appear to have
similar lattice artifacts as asqtad simulations at lattice spacing
$\frac{2}{3}a$, leading to substantial savings in simulation costs.
This program is continuing with 
computations at different light sea-quark masses, so that both the
extrapolation to the continuum limit and the extrapolation to the physical
light-quark mass can be controlled.

\vspace{-0.2in}
\section*{Acknowledgements}
\vspace{-0.25in}
This work was supported by the U.S. Department of Energy grant numbers
DE-FG02-91ER-40628, % Claude
DE-FC02-06ER-41446, % Carleton SciDAC
DE-FG02-91ER-40661, % Steve
DE-FC02-06ER-41443, % Steve SciDAC
DE-FC06-01ER-41437, % Bob SciDAC
DE-FG02-04ER-41298,	% Walter and Doug
DE-FC02-06ER-41439, % Alexei and Doug SciDAC
and by the NSF under grant numbers
PHY05-55235, % CLaude PIF
PHY07-57333, % Carleton
PHY07-03296, % Carleton
PHY05-55243, % Carleton PIF
PHY09-03571, % Carleton PIF
PHY05-55234, % Steve PIF
PHY07-57035, % Bob
OCI-0832315, % Bob
PHY05-55397, % Doug PIF
PHY09-03536, % Doug PIF
and
PHY07-04171. % Jim
Computation for this work was done at
the Texas Advanced Computing Center (TACC),
the National Center for Supercomputing Resources (NCSA)
and
%the Oak Ridge National Laboratory computer center,
the National Institute for Computational Sciences (NICS) under
NSF teragrid allocation TG-MCA93S002.
Computer time at the National Center for Atmospheric Research % Frost
was provided by NSF MRI Grant CNS-0421498, NSF MRI Grant
CNS-0420873, NSF MRI Grant CNS-0420985, NSF sponsorship of the 
National Center for Atmospheric Research, the University of Colorado,
and a grant from the IBM Shared University Research (SUR) program.
Computer time was also provided by the National Energy Resources
Supercomputing Center (NERSC), which is supported by the Office of
Science of the U.S. Department of Energy under Contract No.
DE-AC02-05CH11231.
D.T. would like to thank the University of Colorado for hospitality
during part of this work.
We thank Christine Davies, Alan Gray, Eduardo Follana, and Ron Horgan
for discussions and help in developing and verifying our codes.

% Official frost acknowledgement
%   * Computer time was provided by NSF MRI Grant #CNS-0421498, NSF MRI Grant  |
%|     #CNS-0420873, NSF MRI Grant #CNS-0420985, NSF sponsorship of the         |
%|     National Center for Atmospheric Research, the University of Colorado,    |
%|     and a grant from the IBM Shared University Research (SUR) program.

\appendix

\section{Gauge and fermion actions}\label{app_HISQ_action}

For completeness, we summarize the gauge and fermion actions
in this appendix.

The gauge action is a tadpole-improved \cite{tadpole_improve}
one-loop-Symanzik-improved gauge
action \cite{symanzik_action} including the effects of the quark
loops in the one-loop coefficients \cite{GAUGE_COEFFS}.  The
number of flavors $n_f$ is set to four in these simulations.

This gauge action involves three kinds of loops: the $1\times 1$ loop,
or plaquette $P$, the $2 \times 1$ loop, or rectangle $R$, and the
twisted loop $T$, which traverses paths such as
$+\hat x,+\hat y, +\hat z, -\hat x, -\hat y, -\hat z$.  Then
\BNE S_g = \beta \LP C_P \sum_P \LP 1-\frac{1}{3}\Re\,\Tr(P) \RP
                   + C_R \sum_R \LP 1-\frac{1}{3}\Re\,\Tr(R) \RP
                   + C_T \sum_T \LP 1-\frac{1}{3}\Re\,\Tr(T) \RP  \RP\\\\,
\ENE
where the sums run over all distinct positions and orientations of the loops.
The coefficients are
\BEA
C_P &=& 1.0 \EL
C_R &=& \frac{-1}{20 u_0^2} \LP 1 - \LP 0.6264 - 1.1746 n_f \RP \ln(u_0) \RP \EL
C_T &=& \frac{1}{u_0^2} \LP 0.0433 - 0.0156 n_f \RP \ln(u_0) \ \ \ .\\
\EEA
In this expression the strong coupling constant appears in the form
$\alpha_s = -\ln(u_0)/1.303615$.
With this normalization $\beta = \frac{10}{g^2}$.  We determine the
tadpole coefficient $u_0$ from the average plaquette,
$u_0 = \LP \LL \Re\,\Tr P \RR/3 \RP^{1/4}$.

%    // HISQ coefficients: Ron Horgan private communication
%    loop_coeff[0][0]= 1.0;
%    loop_coeff[1][0]=  -1.00/(20.0*u0*u0) *
%      ( 1.00 - (0.6264 - 1.1746*total_dyn_flavors)*log(u0) );
%    loop_coeff[2][0]=  1.00/(u0*u0) *
%      (0.0433 - 0.0156*total_dyn_flavors) * log(u0);
% also say how u_0 is fixed

% alternatively, 
% rectangle = (1/(10 u_0^2)* ( 1 + 0.4805 + XXX ) alpha_s
% twisted = (1/u_0^2) * (.03325 + XXX) alpha_s
% with alpha_s = -1.303615 log(u_0)

% g_action = (beta/3.0)*imp_gauge_action();
% sum over loops of loop_coeff* (3-Real(trace(loop)))

The fermion factor in the partition function is
\BNE \ln \LP S_f \RP = \prod_f \LP \det\LP 2 \Dslash + 2 m_f \RP \RP^{1/4} \ \ \ .\ENE
The Dirac operator $\Dslash$ is constructed from smeared links.
Two levels of smearing are used, with a projection onto an element of
U(3) after the first smearing.
The fundamental gauge 
links are $U_\mu(x)$, the fat links after a level one fat7 smearing are $V_\mu(x)$,  the
reunitarized links are $W_\mu(x)$, and the fat links after level two asqtad smearing are $X_\mu(x)$.
The first level smeared links $V$ are constructed from the $U$ as a sum
over products of links along paths from $x$ to $x+\hat \mu$, or parallel transports.
\BNE V_\mu(x) = \sum_{paths} \prod_{path} U^{(\dagger)}(path) \ENE
Table~\ref{tab:smearing} gives the coefficients used in the two levels of
smearing.  The nearest neighbor part of $\Dslash$ uses the twice-smeared links $X$ while the third 
nearest neighbor part uses the once-smeared and unitarized links $W$:

\BEA 
\label{eq:dslash_def}
2\Dslash_{x,y} = \sum_\mu 
  && \left\{ \delta_{x+\hat \mu,y} X_\mu(x) - \delta_{x-\hat\mu,y} X_\mu^\dagger(x-\hat\mu) \right\} \EL
+ (1+\epsilon_N) &&\left\{ \delta_{x+3\hat \mu,y} W_\mu(x) W_\mu(x+\hat \mu) W_\mu(x+2\hat\mu) \right. \EL
    && - \left. \delta_{x-3\hat\mu,y} W_\mu^\dagger(x-3\hat\mu) W_\mu^\dagger(x-2\hat\mu)
              W_\mu^\dagger(x-\hat\mu) \right\}
\ \ \ .\\ \EEA

\begin{center}\begin{table}
\caption{\label{tab:smearing}
Paths and coefficients used in smearing the links.  It is understood that all distinct
rotations and reflections of each path are used in the action.  In specifying paths in this table,
directions $\hat x$ and $\hat y$, etc. are different.
The multiplicity is the number of such paths contributing to a single smeared link.
The first block of the table gives the coefficients used in the ``fat7'' smearing
used to construct $V$ from $U$.   The second block gives the coefficients in the ``asqtad+''
smearing used to compute $X$ from the unitarized links $W$, and the final line
is the coefficient of the third nearest neighbor term.   Note that the coefficient
of the ``Lepage'' term that corrects the form factor at small momenta is twice
that of the single smearing asqtad action.
}
\begin{tabular}{|llcl|}
\hline
Name & Path & Multiplicity &Coefficient \\
\hline
Single link & $+\hat x$ & 1 & 1/8 \\
3-staple & $+\hat y\, +\hat x\, -\hat y$ & 6 & $1/16$ \\
5-staple & $+\hat y\, +\hat z\, +\hat x\, -\hat z\,  -\hat y$ & 24 & $1/64$ \\
7-staple & $+\hat y\, +\hat z\, +\hat t\, +\hat x\, -\hat t\,  -\hat z\,  -\hat y$ & 48 & $1/384$ \\
\hline
Single link & $+\hat x$ & 1 & $1/8 + 3/4 + 1/8(1+\epsilon_N)$ \\
3-staple & $+\hat y\, +\hat x\, -\hat y$ & 6 & $1/16$ \\
5-staple & $+\hat y\, +\hat z\, +\hat x\, -\hat z\,  -\hat y$ & 24 & $1/64$ \\
7-staple & $+\hat y\, +\hat z\, +\hat t\, +\hat x\, -\hat t\,  -\hat z\,  -\hat y$ & 48 & $1/384$ \\
``Lepage'' & $+\hat y\,+\hat y\,+\hat x\, -\hat y\, -\hat y$ & 6 & $-1/8$ \\
\hline
``Naik'' & $+\hat x\,+\hat x\,+\hat x$ & 1 & $-1/24(1+\epsilon_N)$  \\
\hline
\end{tabular}
\end{table}\end{center}

In Eq.~(\ref{eq:dslash_def}) and
Table~\ref{tab:smearing}, $\epsilon_N$ is a mass-dependent correction to
the tree-level improvement of the quark dispersion relation, or the ``Naik term.''
This correction is negligible and set to zero for the light and strange quarks.  For the
charm quark we use
\begin{equation}\label{eq:epsilon_N}
\epsilon_N = -\frac{27}{40} (am_c)^2 + \frac{327}{1120} (am_c)^4
       -\frac{15607}{268800} (am_c)^6 - \frac{73697}{3942400} (am_c)^8 \ .
\end{equation}
In this expression $am_c$ is the bare mass in the quark action, and this
formula combines Eqs.~(24) and (26) in Ref.~\cite{hpqcd_hisqprd}.
% note these were Eqs. 20 and 21 in the arXiv version
The numerical values of $\epsilon_N$
used in our simulations are given in Table~\ref{tab:hisqruns}.

% Fermion:
% Fat&
%       ( 1.0/8.0),        /* one link */
%       (-1.0/8.0)*0.5,              /* simple staple */
%       ( 1.0/8.0)*0.25*0.5,         /* displace link in two directions */
%       (-1.0/8.0)*0.125*(1.0/6.0),  /* displace link in three directions */

% project onto U3

% asqtad+
%       (( 1.0/8.0)+(2.0*6.0/16.0)+(1.0/8.0)),        /* one link */
%            /*One link is 1/8 as in fat7 + 2*3/8 for Lepage + 1/8 for Naik */
%       (-1.0/24.0),                 /* Naik */
%       (-1.0/8.0)*0.5,              /* simple staple */
%       ( 1.0/8.0)*0.25*0.5,         /* displace link in two directions */
%       (-1.0/8.0)*0.125*(1.0/6.0),  /* displace link in three directions */
%       (-2.0/16 ),                  /* Correct O(a^2) errors, 2X as much as asqtad  */
%    static Real onelink_mass_renorm_fact = (1.0/8.0)*HISQ_NAIK_EPS;
%    static Real naik_mass_renorm_fact = (-1.0/24.0)*HISQ_NAIK_EPS;

\section{HISQ force calculation details}\label{app_HISQ_force}

\begin{figure}[t]
\hspace{-0.1in}
\rule{0.0in}{0.0in}\vspace{-1.0in}\\
\includegraphics[width=1.0\textwidth]{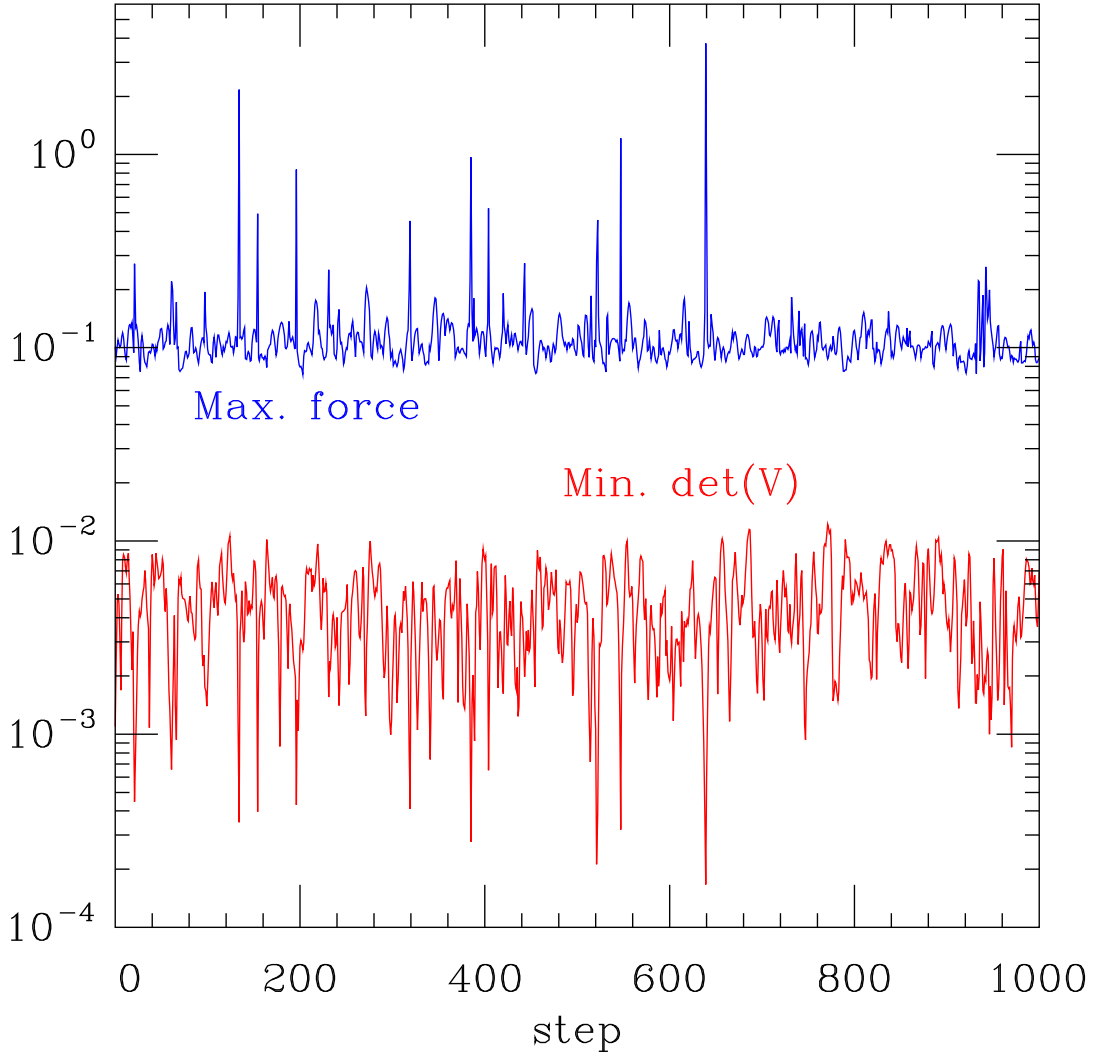}
\rule{0.0in}{0.0in}\vspace{-1.5in}\\
\caption{Time history of the maximum (over lattice sites) magnitude
of the fermion force and the minimum determinant of the fattened
links after the first level of smearing ($V$).
This exploratory
run was done on a $20^3\times64$ lattice at $\beta=6.75$,
$am_l=0.2am_s$, $am_s=0.05$, $am_c=0.6$ and $u_0=0.9$. This
approximately corresponds to the $a\approx 0.12$ fm ensemble in
Table~\ref{tab:hisqruns} at $\beta=6.0$. The difference in
$\beta$ is due to the use of a different gauge action in
the earlier studies.
\label{fig:FF_detV}
}
\end{figure}

Here we summarize the details of the HISQ force calculation. 
For clarity in this and the next appendices we suppress the $x$-dependence and
the direction index in the notation of the links. Most of this
material appeared earlier in Ref.~\cite{milc_hisq}. For the force
calculation we adopted the strategy of Refs.~\cite{Kamleh:2004xk} and~\cite{Wong:2007uz} ,
in which the derivative of the smeared action is calculated by repetitive 
application of the chain rule:
\BNE\label{dSdU_chain}
  \frac{\partial S_f}{\partial U}=
  \frac{\partial S_f}{\partial X}\,\frac{\partial X}{\partial W}\,
  \frac{\partial W}{\partial V}\,\frac{\partial V}{\partial U}\ ,
\ENE
where $S_f$ is the fermion part of the action, $U$ are fundamental gauge 
links, $V$, the fat links after level one fat7 smearing, $W$, the
reunitarized links, and $X$, the fat links after level two asqtad smearing.
In our code, for the parts that involve smearing, we follow the same procedure
as for the asqtad action.  This procedure is described in
Refs.~\cite{ASQTADFORCE1} and \cite{ASQTADFORCE2},
so we do not repeat it here. The algorithm for the reunitarization part is 
detailed below.

We have chosen to project links to U(3), rather than SU(3) as in the
original HPQCD/UKQCD formulation, for two reasons:
\begin{enumerate}
  \item SU(3) projection requires calculation of the third root
  of the determinant, which involves a phase that can initially be
  restricted to, \textit{e.g.}, the interval $[-\pi/3,\pi/3)$. However,
  during the molecular dynamics evolution, this phase has to evolve continuously
  (to prevent the appearance of $\delta$ function-like forces) and may cross
  into $[\pi/3,2\pi/3)$ interval, and so on. Thus, SU(3) projection
  requires tracking the evolution of the phase for each link during molecular dynamics.
  \item For the U(3) group, different methods of projection yield the same
  answer for the projected link, $W$.
\end{enumerate}

For instance, the default method in our code is polar projection: 
one builds a Hermitian matrix
\BNE
  Q=V^\dagger V
\ENE
and then
\BNE\label{WeqVQ_polar}
  W=VQ^{-1/2}
\ENE
belongs to U(3), \textit{i.e.},
\BNE
  W^\dagger W=\left(Q^{-1/2}\right)^\dagger V^\dagger V Q^{-1/2}=
  Q^{-1/2}QQ^{-1/2}=1.
\ENE
It is important that closed form expressions for $Q^{-1/2}$ can be
derived~\cite{Hasenfratz:2007rf} and, thus, the whole procedure can be
implemented analytically.

One may expect, given that $Q^{-1/2}$ is a singular operation,
that when one of the eigenvalues of $Q$ is close to 0, the numerical
accuracy in evaluation of $W$ becomes poor. In fact, in simulations
one occasionally encounters large deviations from unitarity:
\BNE
  \left|W^\dagger W-1\right|\sim O(1).
\ENE
Such situations are rare, but the contribution of (systematic) errors of
this kind can be large. Therefore, we also implemented the singular value
decomposition (SVD) algorithm, which is slower, but is used only in
exceptional cases. We decompose:
\BNE\label{VeqASigmaB}
  V=A\Sigma B^\dagger,
\ENE
$A, B \in{\rm U(3)}$ and $\Sigma$ is a positive, diagonal matrix. 
(The values on the
diagonal are called the singular values of $V$.) Then we have, simply,
\BNE\label{WeqAB_svd}
  W=AB^\dagger.
\ENE
It is easy to see that (\ref{WeqVQ_polar}) gives the same result as
(\ref{WeqAB_svd}):
\BEA
  Q&=&V^\dagger V=B\Sigma A^\dagger A\Sigma B^\dagger=
    B\Sigma \Sigma B^\dagger=B\Sigma B^\dagger B\Sigma B^\dagger=
    \left(B\Sigma B^\dagger\right)^2\ ,\nonumber\\
  Q^{-1/2}&=&\left(B\Sigma B^\dagger\right)^{-1}=
    \left(B^\dagger\right)^{-1}\Sigma^{-1}B^{-1}=B\Sigma^{-1}B^\dagger\ ,
    \nonumber\\
  W&=&A\Sigma B^\dagger B\Sigma^{-1}B^\dagger=AB^\dagger\ ,
\EEA
as it should be. The SVD algorithm of Golub and Reinsch~\cite{svd_1970}
is numerically stable even in the case of exactly zero eigenvalues.

Another popular projection algorithm is ``trace maximization'' \cite{trace_max}:
find $W\in {\rm U(3)}$ such that it maximizes
\BNE
  \Omega = {\rm Re}{\rm Tr} \left\{ V^\dagger W\right\}\ .
\ENE
Let us again use SVD on $V$:
\BNE\label{Vmax}
  V=A\Sigma B^\dagger\ \ \ \Rightarrow\ \ \ 
  \Omega = {\rm Re}{\rm Tr}\left\{B\Sigma A^\dagger W\right\}=
  {\rm Re}{\rm Tr}\left\{\Sigma A^\dagger W B\right\}\ .
\ENE
Since $\Sigma$ is positive, clearly $\Omega$ is maximized when
\BNE
  A^\dagger W B=1\ \ \ \Rightarrow\ \ \ W=AB^\dagger
\ENE
and we arrive at (\ref{WeqAB_svd}) again. In the SU(3) case an extra phase
present in (\ref{Vmax}) would lead to $W$ different from (\ref{WeqAB_svd}).
To summarize, we use the polar projection (\ref{WeqVQ_polar}) replaced
by SVD if small eigenvalues of $Q$ are encountered.

Appendix~\ref{app_algebra} gives the details of calculation of $Q^{-1/2}$ and its derivative 
$\partial Q^{-1/2}/\partial V$ within the approach of 
Refs.~\cite{Morningstar:2003gk} and~\cite{Hasenfratz:2007rf} 
based on the Cayley-Hamilton theorem.

Let us now turn to the calculation of the force. During the molecular dynamics evolution, one
encounters (more often on coarser ensembles) matrices $V$ that have small
eigenvalues. Let us consider the U(1) group for simplicity. Then $V$ is
just an arbitrary complex number $V=re^{i\theta}$. The projection
onto U(1) is $W=e^{i\theta}$. The derivative that enters the force
calculation is
\BNE
  \frac{\partial W}{\partial V}\equiv\left(\frac{\partial W}{\partial V}
  \right)_{V^\dagger}=\frac{\partial(W,V^\dagger)}{\partial(V,V^\dagger)}=
  \frac{\partial(W,V^\dagger)}{\partial(r,\theta)}\,
  \frac{\partial(r,\theta)}{\partial(V,V^\dagger)}=\frac{1}{2r},
\ENE
Thus the derivative is inversely proportional to the magnitude of
$V$, or, in the U(3) case, to the smallest singular value of $V$ (or eigenvalue of 
$Q$), which is not protected from being zero. Thus, on rare occasions one has
to deal with exceptionally large forces that give large contributions to the
action, but originate from a single link. In Fig.~\ref{fig:FF_detV} we show
the evolution of the minimal $\det|V|$ over the lattice and maximal
value of the norm of the fermion force on a logarithmic scale. One can
easily see the correlation: the lower $\det|V|$, the higher the force.

To circumvent this problem we 
introduce a ``cutoff'' in the force calculation by replacing:
\BNE\label{WeqVQdelta}
  W=VQ^{-1/2}\ \ \ \rightarrow\ \ \ \ W=V(Q+\delta I)^{-1/2},
\ENE
where $I$ is the unit matrix,
whenever the smallest eigenvalue of $Q$ is less than $\delta$.
In the ensembles used in this paper we set $\delta=5\times 10^{-5}$.
In tuning $\delta$, we weigh two competing issues: The value of $\delta$ should be
large enough to suppress an exceptional contribution from a link, but
small enough not to modify too many forces on the lattice. If $\delta$ is too large,
the evolution will be smooth, but the fluctuation of the action
will be large, usually leading to rejection of the trajectory.
Note that
we modify $W$ only in the force calculation, and we use the original 
Eq.~(\ref{WeqVQ_polar}), or Eq.~(\ref{WeqAB_svd}) for nearly 
singular matrices, to calculate the action at the accept/reject step.
That is, the modification (\ref{WeqVQdelta}) amounts to
using a different guiding Hamiltonian during the evolution, while the
Metropolis step insures the desired distribution.

\section{Algebra for reunitarized links and their derivatives}\label{app_algebra}

To make this presentation self-contained we include
Eqs.~(\ref{Q12})-(\ref{Cij_elements}) from Hasenfratz, Hoffmann and Schaefer
\cite{Hasenfratz:2007rf}, preserving the notation 
of the original.

The inverse square root of a non-singular matrix $Q$ 
entering Eq.~(\ref{WeqVQ_polar}) is given
by the Cayley-Hamilton theorem as a polynomial of $Q$:
\begin{equation}\label{Q12}
  Q^{-1/2}=f_0+f_1Q+f_2Q^2.
\end{equation}
Since $Q$ is Hermitian, it has nonnegative eigenvalues that 
can be found by solving the characteristic equation
\begin{equation}\label{g_i_cubic}
  g^3-c_0g^2-\left(c_1-\frac{1}{2}c_0^2\right)g
  -\left(c_2-c_0c_1+\frac{1}{6}c_0^3\right)=0,
\end{equation}
where
\begin{equation}\label{c_n_def}
c_{n}=\frac{1}{n+1}{\textrm{tr}}\, Q^{n+1}\,,\,\,\,\,\,\,
n=0,1,2\,.
\end{equation}
The solution of the cubic equation (\ref{g_i_cubic}) is
\begin{equation}\label{g_n_final}
  g_{n}=\frac{c_{0}}{3}
  +2\sqrt{S}\,\cos\left(\frac\theta3+(n-1)\frac{2\pi}{3}\right)\,,
  \,\,\,\,\,n=0,1,2,
\end{equation}
where
\begin{equation}
  S=\frac{c_{1}}{3}-\frac{c_{0}^{2}}{18}\,,\,\,\,\,\,
  R=\frac{c_{2}}{2}-\frac{c_{0}c_{1}}{3}+\frac{c_{0}^{3}}{27}\,,
  \,\,\,\,\,
  \theta=\arccos\left(\frac{R}{S^{3/2}}\right)\,.
\end{equation}
It is convenient to define the symmetric polynomials 
of the square roots of the eigenvalues 
\begin{eqnarray}
  u&=&\sqrt{g_{0}}+\sqrt{g_{1}}+\sqrt{g_{2}}\,,\nonumber\\
  v&=&\sqrt{g_{0}g_{1}}+\sqrt{g_{0}g_{2}}+\sqrt{g_{1}g_{2}}\,,\nonumber\\
  w&=&\sqrt{g_{0}g_{1}g_{2}}\,.
\end{eqnarray}
In the diagonalized form the expression (\ref{Q12}) can be rewritten
as an equation for $f_i$
\begin{equation}\label{eq_f_i}
  \left(\begin{array}{ccc}
  1 & g_{0} & g_{0}^{2}\\
  1 & g_{1} & g_{1}^{2}\\
  1 & g_{2} & g_{2}^{2}\end{array}\right)\left(\begin{array}{c}
  f_{0}\\
  f_{1}\\
  f_{2}\end{array}\right)=
  \left(\begin{array}{c}
  g_{0}^{-1/2}\\
  g_{1}^{-1/2}\\
  g_{2}^{-1/2}\end{array}\right)
\end{equation}
which has the solution
\begin{eqnarray}
  f_{0} & = & \frac{-w(u^{2}+v)+uv^{2}}{w(uv-w)}\,,\nonumber \\
  f_{1} & = & \frac{-w-u^{3}+2uv}{w(uv-w)}\,,\nonumber\\
  f_{2} & = & \frac{u}{w(uv-w)}\,.\label{f_i_def}
\end{eqnarray}
The derivative $\partial f_i/\partial c_j$ can be written as
\begin{equation}\label{Bij}
  B_{ij}\equiv\frac{\partial f_i}{\partial c_j}=
  \sum_{k=0}^{2}\frac{\partial f_i}{\partial g_k}
  \frac{\partial g_k}{\partial c_j}\,.
\end{equation}
After rescaling~(\ref{Bij}) by the common denominator
\begin{equation}
  C_{ij}\equiv dB_{ij}\,,\,\,\,\,\,\,\,
  d=2w^{3}(uv-w)^{3}
\end{equation}
a closed-form expression for the symmetric matrix $C_{ij}$
has been derived in Ref.~\cite{Hasenfratz:2007rf}
\begin{eqnarray}
  C_{00} & = &-w^{3}u^{6}+3vw^{3}u^{4}+3v^{4}wu^{4}-
    v^{6}u^{3}-4w^{4}u^{3}-12v^{3}w^{2}u^{3}\nonumber\\
    &  & +16v^{2}w^{3}u^{2}+3v^{5}wu^{2}-
    8vw^{4}u-3v^{4}w^{2}u+w^{5}+v^{3}w^{3}\,,\nonumber\\
  C_{01} & = & -w^{2}u^{7}-v^{2}wu^{6}+v^{4}u^{5}
    +6vw^{2}u^{5}-5w^{3}u^{4}-v^{3}wu^{4}-2v^{5}u^{3}\nonumber\\
    &  & -6v^{2}w^{2}u^{3}+10vw^{3}u^{2}+
    6v^{4}wu^{2}-3w^{4}u-6v^{3}w^{2}u+2v^{2}w^{3}\,,\nonumber\\
  C_{02} & = & w^{2}u^{5}+v^{2}wu^{4}-v^{4}u^{3}
    -4vw^{2}u^{3}+4w^{3}u^{2}+3v^{3}wu^{2}-3v^{2}w^{2}u+vw^{3}\,,\nonumber\\
  C_{11} & = & -wu^{8}-v^{2}u^{7}+7vwu^{6}+4v^{3}u^{5}-5w^{2}u^{5}
    -16v^{2}wu^{4}-4v^{4}u^{3}+16vw^{2}u^{3}\nonumber\\
    &  & -3w^{3}u^{2}+12v^{3}wu^{2}-12v^{2}w^{2}u+3vw^{3}\,,\nonumber\\
  C_{12} & = & wu^{6}+v^{2}u^{5}-5vwu^{4}-2v^{3}u^{3}
    +4w^{2}u^{3}+6v^{2}wu^{2}-6vw^{2}u+w^{3}\,,\nonumber\\
  C_{22} & = & -wu^{4}-v^{2}u^{3}+3vwu^{2}-3w^{2}u\,.\label{Cij_elements}
\end{eqnarray}
In the following, differentiation with respect to $V$ at fixed $V^\dagger$
is always assumed. We use explicit color indices to show how
different contractions and direct products of matrices are built.

The derivatives that enter in the calculation of the fermion force
are
\begin{eqnarray}
  \frac{\partial W_{ij}}{\partial V_{kl}}&=&
  \frac{\partial (V_{im}(Q^{-1/2})_{mj})}{\partial V_{kl}}=
  \delta_{ik}(Q^{-1/2})_{lj}+V_{im}
  \frac{\partial (Q^{-1/2})_{mj}}{\partial V_{kl}}\,,\label{dWdV}\\
  \frac{\partial W^\dagger_{ij}}{\partial V_{kl}}&=&
  \frac{\partial ((Q^{-1/2})_{im}V^\dagger_{mj})}{\partial V_{kl}}=
  \frac{\partial (Q^{-1/2})_{im}}{\partial V_{kl}}V^\dagger_{mj}\,.
  \label{dWdagdV}
\end{eqnarray}
Also,
\begin{equation}
  \frac{\partial Q_{ij}}{\partial V_{kl}}=V^\dagger_{ik}\delta_{lj}.
\end{equation}
The central component of the calculation is
\begin{eqnarray}
  \frac{\partial (Q^{-1/2})_{ij}}{\partial Q_{pq}}&=&
  \frac{\partial}{\partial Q_{pq}}
  \left(
  f_0\delta_{ij}+f_1Q_{ij}+f_2(Q^2)_{ij}
  \right)\nonumber\\
  &=&\frac{\partial f_0}{\partial Q_{pq}}\delta_{ij}
  +\frac{\partial f_1}{\partial Q_{pq}}Q_{ij}
  +f_1\delta_{ip}\delta_{qj}
  +\frac{\partial f_2}{\partial Q_{pq}}(Q^2)_{ij}
  +f_2\left(\delta_{ip}Q_{qj}+Q_{ip}\delta_{qj}\right)\,.\label{dQ12dQ}
\end{eqnarray}
From the definition~(\ref{c_n_def}) it follows that
$\partial c_n/\partial Q_{pq}=(Q^n)_{pq}$. Then
\begin{equation}
  \frac{\partial f_k}{\partial Q_{pq}}=
  \sum_{n=0}^2\frac{\partial f_k}{\partial c_n}
  \frac{\partial c_n}{\partial Q_{pq}}=
  \sum_{n=0}^2 B_{kn}(Q^n)_{pq}\,.
\end{equation}
We define
\begin{eqnarray}
  P_{qp}\equiv\frac{\partial f_0}{\partial Q_{pq}}
  =B_{00}\delta_{qp}+B_{01}Q_{qp}+B_{02}(Q^2)_{qp}\,,\label{Pqp_def}\\
  R_{qp}\equiv\frac{\partial f_1}{\partial Q_{pq}}
  =B_{10}\delta_{qp}+B_{11}Q_{qp}+B_{12}(Q^2)_{qp}\,,\label{Rqp_def}\\
  S_{qp}\equiv\frac{\partial f_2}{\partial Q_{pq}}
  =B_{20}\delta_{qp}+B_{21}Q_{qp}+B_{22}(Q^2)_{qp}\,.\label{Sqp_def}
\end{eqnarray}
Substituting~(\ref{Pqp_def}), (\ref{Rqp_def}) and (\ref{Sqp_def}) into
(\ref{dQ12dQ}) and Eq.~(\ref{dQ12dQ}) in (\ref{dWdV}) and (\ref{dWdagdV})
we obtain the final result
\begin{eqnarray}
  \frac{\partial W_{ij}}{\partial V_{kl}}&=&
  \delta_{ik}(Q^{-1/2})_{lj}+\left[f_1(VV^\dagger)_{ik}
  +f_2(VQV^\dagger)_{ik}\right]\delta_{lj}
  +f_2(VV^\dagger)_{ik}Q_{lj}\nonumber\\
  &+&V_{ij}(PV^\dagger)_{lk}+(VQ)_{ij}(RV^\dagger)_{lk}
  +(VQ^2)_{ij}(SV^\dagger)_{lk}\,,\label{dWdVfinal}\\
  \frac{\partial W^\dagger_{ij}}{\partial V_{kl}}&=&
  \left[f_1V^\dagger_{ik}+f_2(QV^\dagger)_{ik}\right]
  V^\dagger_{lj}+f_2V^\dagger_{ik}(QV^\dagger)_{lj}\nonumber\\
  &+&V^\dagger_{ij}(PV^\dagger)_{lk}+
  (QV^\dagger)_{ij}(RV^\dagger)_{lk}+
  (Q^2V^\dagger)_{ij}(SV^\dagger)_{lk}\ \ \ .
  \label{dWdagdVfinal}
\end{eqnarray}

The calculation of the fermion force from the reunitarized links 
proceeds as follows:
\begin{enumerate}
  \item The eigenvalues of the Hermitian matrix $Q$ 
  are calculated with Eq.~(\ref{g_n_final}).
  \item $\det|Q|$ is
  compared with the product $g_0g_1g_2$. If the relative error
  is larger than $10^{-8}$ or any eigenvalue is
  smaller than $10^{-8}$ the singular value decomposition of
  $V$ is performed and the eigenvalues are set to
  \begin{equation}
    g_i=\sigma_i^2\,,\,\,\,\,\,\,\,i=0,1,2,
  \end{equation}
  where $\sigma_i$ are the diagonal elements of the matrix
  $\Sigma$ in Eq.~(\ref{VeqASigmaB}). 
  \item Additionally, if any
  of the eigenvalues is smaller than 
  (an adjustable parameter) $\delta=5\times10^{-5}$
  the eigenvalues are modified to
  \begin{equation}
    g_i\,\,\,\rightarrow\,\,\, g_i+\delta.
  \end{equation}
  (This corresponds to the force ``cutoff'' in Eq.~(\ref{WeqVQdelta}).)
  \item With these eigenvalues the coefficients $f_i$ and 
  the elements $B_{ij}$ are calculated from Eq.~(\ref{f_i_def})
  and (\ref{Cij_elements}).
  \item Finally, the force is calculated from Eq.~(\ref{dWdVfinal})
  and (\ref{dWdagdVfinal}).
\end{enumerate}

In the MILC code we have also implemented two other methods 
for calculating $Q^{-1/2}$ and its derivative:
\begin{enumerate}
  \item a rational function approximation,
  \item an iterative evaluation of $Q^{-1/2}$ 
  with the derivative replaced by finite difference.
\end{enumerate}
We found that the analytic evaluation via Eq.~(\ref{dWdVfinal}) and
(\ref{dWdagdVfinal}) is superior to the other methods
due to its higher precision and speed.

\section{Treatment of the HISQ charm quark}\label{app_charm}

\begin{figure}[t]
\hspace{-0.1in}
\rule{0.0in}{0.0in}\vspace{-1.0in}\\
\includegraphics[width=1.0\textwidth]{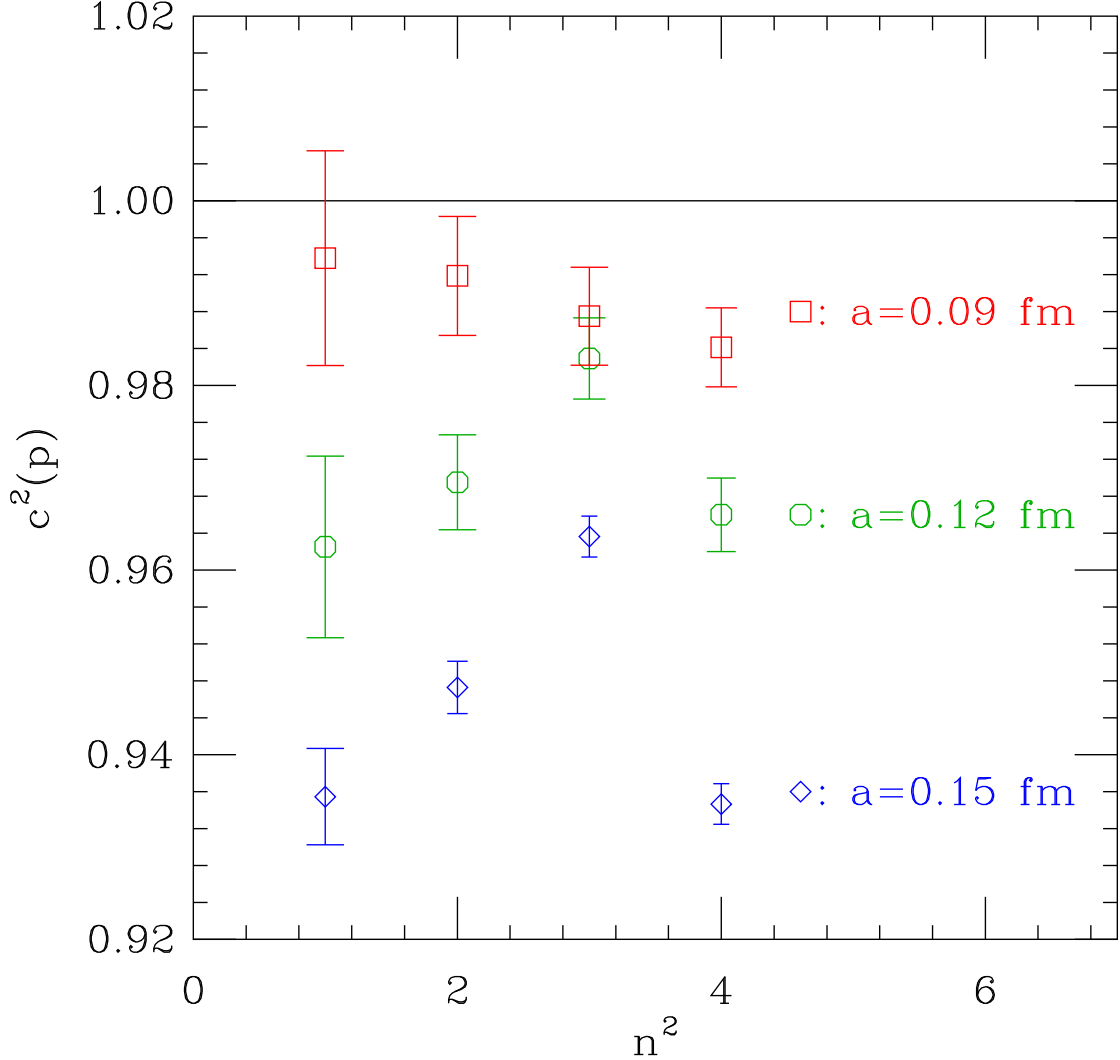}
\rule{0.0in}{0.0in}\vspace{-1.5in}\\
\caption{The speed of light for the $\eta_c$ meson calculated at 
several values of momenta. 
\label{fig:disp_eta_c}
}
\end{figure}

The tree-level discretization errors are $O((ap_\mu)^4)$ and are 
negligible for light quarks. However, 
at the lattice spacings listed in Table~\ref{tab:hisqruns}, the charm quark mass 
is in the range $am_c\sim 0.4-0.8$ and therefore the discretization
errors are larger. The leading tree-level $O((am_c)^4)$ error can be removed
by retuning the coefficient of the third-nearest-neighbor (Naik) 
term~\cite{hpqcd_hisqprd}, using the expansion in Eq.~(\ref{eq:epsilon_N}).
%\begin{equation}\label{eq:epsilon_N}
%\epsilon_N = -\frac{27}{40} (am_c)^2 + \frac{327}{1120} (am_c)^4
       %-\frac{15607}{268800} (am_c)^6 - \frac{73697}{3942400} (am_c)^8 \ .
%\end{equation}
%In this expression $am_c$ is the bare mass in the quark action, and this
%formula combines Eqs.~(20) and (21) in Ref.~\cite{hpqcd_hisqprd}.
As can be seen from Table~\ref{tab:hisqruns}, at the finest lattice,
$a\approx 0.09$ fm, it is quite small, $\epsilon_N=-0.120471$.
%To examine the effect of the correction (\ref{eq:epsilon_N}) we computed
% DT since we don't do this with and without the correction, we don't
% really examine the effect of the correction
The effect of the correction in Eq.~(\ref{eq:epsilon_N}) has been studied
in Ref.~\cite{hpqcd_hisqprd}.
To check the quality of charm quark physics in our ensembles, we computed
the speed of light for the $\eta_c$ meson by calculating its propagator
at several non-zero momenta. The result is shown in Fig.~\ref{fig:disp_eta_c},
where
\begin{equation}\label{c2_eta_c}
  c^2(p)=\frac{E^2(p)-E^2(0)}{p^2}
\end{equation}
and the momenta are rescaled by the lattice size $L_s$
\begin{equation}
  n^2=p^2\,\left(\frac{L_s}{2\pi}\right)^2.
\end{equation}
For the finest $a\approx 0.09$ fm ensemble the error in the dispersion relation
is below 2\%.
As expected, these $\eta_c$ dispersion relations are very similar to those found for
HISQ valence quarks on asqtad sea quarks, shown in Table V of Ref.~\cite{hpqcd_hisqprd}.

For dynamical simulations, the mass-dependent correction to the Naik term 
requires the use of different sets of smeared links for light quarks and
the charm quark. Since the difference in the Naik term enters at the second level
of smearing, it is advantageous to regroup the force calculation as described
in the following. Let $X^{(0)}$ denote the fat links after level two asqtad
smearing for the light quarks, for which $\epsilon_N$ is set to zero,
and $X^{(c)}$ denote the fat links for the charm quark.
Then the fat links for the charm quark can be written as
\begin{equation}
  X^{(c)}=X^{(0)}+\epsilon_N\Delta X\,,
\end{equation}
where $\Delta X$ contains only one-link and three-link paths. 
(This can be easily seen from Table~\ref{tab:smearing}.)
The derivative is
\BNE\label{dXdW_dDeltaX}
  \frac{\partial X^{(c)}}{\partial W}=
  \frac{\partial X^{(0)}}{\partial W}+
  \epsilon_N\frac{\partial \Delta X}{\partial W}.
\ENE
The fermion force in Eq.~(\ref{dSdU_chain}) contains contributions from the
light ($u$, $d$ and $s$) quarks, and from the charm quark.
\BNE\label{dSdU_chain_Xe}
  \frac{\partial S_f}{\partial U}=
  \frac{\partial S_f}{\partial X^{(0)}}\,\frac{\partial X^{(0)}}{\partial W}\,
  \frac{\partial W}{\partial V}\,\frac{\partial V}{\partial U}+
  \frac{\partial S_f}{\partial X^{(c)}}\,
  \frac{\partial X^{(c)}}{\partial W}\,
  \frac{\partial W}{\partial V}\,\frac{\partial V}{\partial U}\ .
\ENE
The calculation of the force for multiply smeared
actions proceeds from the last level of smearing to the first one. Therefore,
operations with $X^{(0)}$ and $X^{(c)}$ links are done first and can be
combined before the reunitarization part:
\begin{eqnarray}
\label{dSdU_chain_Xe2}
  \frac{\partial S_f}{\partial U}&=&
  \left(
  \frac{\partial S_f}{\partial X^{(0)}}\,\frac{\partial X^{(0)}}{\partial W}\,
  +\frac{\partial S_f}{\partial X^{(c)}}\,
  \frac{\partial X^{(c)}}{\partial W}\right)\,
  \frac{\partial W}{\partial V}\,\frac{\partial V}{\partial U}\nonumber\\
  &=&  \left(
  \left(
  \frac{\partial S_f}{\partial X^{(0)}}+
  \frac{\partial S_f}{\partial X^{(c)}}\right)
  \,\frac{\partial X^{(0)}}{\partial W}
  +\epsilon_N\frac{\partial S_f}{\partial X^{(c)}}\,
  \frac{\partial \Delta X}{\partial W}\right)\,
  \frac{\partial W}{\partial V}\,\frac{\partial V}{\partial U}\,.
\end{eqnarray}
After the $\partial\Delta X/\partial W$ contribution is separated,
the number of operations needed for the HISQ fermion force is reduced to slightly 
more than twice the number needed for the asqtad fermion force.
This is because the most time-consuming part of the calculation
is related to 3-, 5- and 7-staple paths that have high multiplicity.
In our final form (\ref{dSdU_chain_Xe2}) they are present only in 
$\partial X^{(0)}/\partial W$ and $\partial V/\partial U$.

\end{document}